\documentclass[oneside,onecolumn]{article}

\usepackage{blindtext} % Package to generate dummy text throughout this template
\usepackage{epsfig}
\usepackage{supertabular,lscape,epsfig}
\usepackage{subfloat}
\usepackage{xcolor}
\usepackage{lineno}
\usepackage{amssymb}
\usepackage{amsmath}
\usepackage{mathtools}
\usepackage{fonttable}
\usepackage{letltxmacro}
\usepackage{subfigure}
\usepackage{cases}
\usepackage{graphicx}
\usepackage{txfonts}
\usepackage[T1]{fontenc}
\usepackage{lipsum}
\usepackage{natbib}
\usepackage[ps2pdf,%
a4paper=true,%
breaklinks=true,%
colorlinks=true,%
pdfauthor={First Author et al.},%
pdftitle={}%
]{hyperref}
\usepackage[sc]{mathpazo} % Use the Palatino font
\usepackage[T1]{fontenc} % Use 8-bit encoding that has 256 glyphs
\usepackage{microtype} % Slightly tweak font spacing for aesthetics

\usepackage[english]{babel} % Language hyphenation and typographical rules

\usepackage[hmarginratio=1:1,top=32mm,columnsep=20pt]{geometry} % Document margins
\usepackage[hang, small,labelfont=bf,up,textfont=it,up]{caption} % Custom captions under/above floats in tables or figures
\usepackage{booktabs} % Horizontal rules in tables

\usepackage{lettrine} % The lettrine is the first enlarged letter at the beginning of the text

\usepackage{enumitem} % Customized lists
\setlist[itemize]{noitemsep} % Make itemize lists more compact

\usepackage{abstract} % Allows abstract customization
\usepackage{titlesec} % Allows customization of titles
\renewcommand\thesection{\Roman{section}} % Roman numerals for the sections
\renewcommand\thesubsection{\roman{subsection}} % roman numerals for subsections
\titleformat{\section}[block]{\large\scshape\centering}{\thesection.}{1em}{} % Change the look of the section titles
\titleformat{\subsection}[block]{\large}{\thesubsection.}{1em}{} % Change the look of the section titles
\usepackage{fancyhdr} % Headers and footers
\usepackage{titling} % Customizing the title section
\usepackage{hyperref} % For hyperlinks in the PDF
\pretitle{\begin{center}\Huge\bfseries} % Article title formatting
\posttitle{\end{center}} % Article title closing formatting
\title{Temperature Analysis of Flaring (AR11283) and non-Flaring (AR12194) Coronal Loops}
\author{\textsc{N.~ Fathalian$^1$,~~~S.~S.~Hosseini~Rad$^2$,~~~N.~Alipour$^2$, H.~Safari$^2$}
\\$^1$Department of Physics, Payame Noor University (PNU), 19395-3697, Tehran, Iran.
\\$^2$Department of Physics, Faculty of Science, University of Zanjan, 45195-313, Zanjan, Iran.
\\e-mail: narges$\_$fathalian@alum.sharif.edu
}
\date{\today} % Leave empty to omit a date
\renewcommand{\maketitlehookd}{%
\begin{abstract}
Here, we study the temperature structure of flaring and non-flaring coronal loops, using extracted loops from images taken in six extreme ultraviolet (EUV) channels recorded by Atmospheric Imaging Assembly (AIA)/ Solar Dynamic Observatory (SDO). We use data for loops of X2.1-class-flaring active region (AR11283) during 22:10UT till 23:00UT, on 2011, September 6; and non-flaring active region (AR12194) during 08:00:00UT till 09:00:00UT on 2014, October 26. By using spatially-synthesized Gaussian DEM forward-fitting method, we calculate the peak temperatures for each strip of the loops. We apply the Lomb-Scargle method to compute the oscillations periods for the temperature series of each strip. The periods of the temperature oscillations for the flaring loops are ranged from 7 min to 28.4 min. These temperature oscillations show very close behavior to the slow-mode oscillation. We observe that the temperature oscillations in the flaring loops are started at least around 10 minutes before the transverse oscillations and continue for a long time duration even after the transverse oscillations are ended. The temperature amplitudes are increased at the flaring time (during 20 min) in the flaring loops.
The periods of the temperatures obtained for the non-flaring loops are ranged from 8.5 min to 30 min,but their significances are less (below 0.5) in comparison with the flaring ones (near to one). Hence the detected temperature periods for the non-flaring loops' strips are less probable in comparison with the flaring ones, and maybe they are just fluctuations.
 Based on our confined observations, it seems that the flaring loops' periods show more diversity and their temperatures have wider ranges of variation than the non-flaring ones. More accurate commentary in this respect requires more extensive statistical research and broader observations.
\\{Coronal Loops,Temperature Analysis, Temperature Oscillations,Flaring and non-Flaring Active Regions}
\end{abstract}
}
\begin{document}
%%%%%%%%%%%%%%%%%%%%%%%%%%%%%%%%%%%%%%%%%%%%%%%%%%%%%%%%%%%%%%%%%%%%%%%%%%%%%
%% Frontmatter
%\begin{frontmatter}
\maketitle

%\author{M.~ A.~ M~o~r~a~d~h~a~s~e~l~i$^1$,~~~M.~ J~a~v~a~h~e~r~i~a~n$^2$,~~~N.~ F~a~t~h~a~l~i~a~n$^3$,~~and~~~H.~ S~a~f~a~r~i$^1$}
%\fntext[footnote1]{Department of Physics, Faculty of Science, University of Zanjan, 45195-313, Zanjan, Iran.}
%\fntext[footnote2]{Research Institute for Astronomy and Astrophysics of Maragha $(RIAAM)$, University of Maragheh, 55136-553, Maragheh, Iran.
%e-mail:~~javaherian@maragheh.ac.ir}
%\fntext[footnote1]{Department of Physics, Payame Noor University$(PNU)$, 19395-3697, Tehran, Iran}

%\end{frontmatter}

\parindent=0.5 cm

%%%%%%%%%%%%%%%%%%%%%%%%%%%%%%%%%%%%%%%%%%%%%%%%%%%%%%%%%%%%%%%%%%%%%%%%%%%%%
%% Main text
\section{Introduction}\label{sect:intro}

Analyzing the thermal structure of coronal loops is of considerable interest, especially as these magnetic loops have an essential role in heating the solar chromosphere and corona. Such analysis can help to describe how the process of solar flaring is correlated with the loop's thermal structure. 

Detections of coronal waves have a historical preview and have been reported for several times (e.g., \citet{ref:Aschwanden1999}; \citet{ref:Nakariakov1999};\citet{ref:Wang2003}; \citet{ref:Wang2004}; \citet{ref:Berghmans1999}; \citet{ref:DeMoortel2000}, \citet{ref:Verwichte2004}, \citet{ref:DeMoortel2007}, \citet{ref:Ballai2011}). Coronal seismology and MHD waves have been reviewed widely by \citet{ref:DeMoortel2005}, \citet{ref:Nakariakov2005}, \citet{ref:Aschwanden2006}, \citet{ref:Banerjee2007} and \citet{ref:DeMoortel2012}. Along with the development of the observations, transverse and longitudinal oscillations have also been studied theoretically (e.g., \citet{ref:Gruszecki2006}, \citet{ref:Pascoe2007}, \citet{ref:Fathalian2009}; \citet{ref:Luna2010};  \citet{ref:Fathalian2010}. Coronal seismology techniques help to elicit the information from observations of oscillatory phenomena and the results to be interpreted by using theoretical models (see for e.g., \citet{ref:Roberts1984}; \citet{ref:Goossens1992}). Oscillatory patterns and processes which happen during solar flares, were interesting and subject of investigations from different approaches (e.g., \citet{ref:Nakariakov2010}, \citet{ref:Nistico2013}, \citet{ref:Anfinogentov2013}, \citet{ref:Hindman2014}, \citet{ref:Russell2015}). As we know the transverse loops oscillations usually occur in response to a close filament or flare (\citet{ref:Wills-Davey1999}). 

Rapidly decaying long-period oscillations are mostly interpreted as global (or fundamental mode) standing slow magnetoacoustic waves (reviewed by \citet{Liu2014}, and \citet{ref:Wang2011}, also see \citet{Ofman2002}, and for slow-mode observed in fan-loops see \citet{ref:Pant2017}). They often occur in hot coronal loops of active regions, associated with tiny (or micro-) flares.Increasing evidence has suggested that the harmonic type of decaying pulsations detected in intensity plots of solar and stellar flares are possibly caused by standing slow-mode waves (see reviews by \citet{ref:VanDoorsselaere2016}, and \citet{ref:McLaughlin2018}).Excitation, propagation, and damping mechanisms of slow-mode waves have been studied theoretically (e.g., \citet{ref:Wang2007}; \citet{ref:Wang2015}; \citet{ref:Jess2016}; \citet{ref:Nakariakov2017}; \citet{ref:Nistico2017}; \citet{ref:Kolotkov2019}; \citet{ref:Krishna2019}; \citet{ref:Reale2019}; \citet{ref:WangO2019}). To have a complete overview of slow-mode magnetoacoustic waves in coronal loops see the review by \citet{ref:Wang2021}. 

Investigating and comparing the thermal structures and oscillations of coronal loops in loops of flaring and non-flaring active regions could help us in better understanding the loops' material oscillations and the flare impact on them. Several different methods have been developed to investigate the thermal structure of the coronal loops and loop strands. The thermal stability of the coronal loops was the subject of research, done by \citet{ref:Habbal1979} (and references cited therein). \citet{ref:McClymont1985} stated that a pressure fluctuation must assist asymmetric coronal temperature perturbation. They concluded that coronal loops are impartially stable in the case of uniform heating. \citet{ref:VanDoorsselaere2011} used spectroscopic line ratios to obtain the required temperature (via CHIANTI code) and estimated the adiabatic index of the corona. The dependence of coronal loop temperature on loop length and magnetic field strength is also a favorite topic. For instance, \citet{ref:Dahlburg2018} probed the temperature properties of solar coronal loops over a wide range of lengths and magnetic field strengths via numerical simulations and observed a very high correlation between magnetic field strength and a maximum of the temperature. The effect of temperature inhomogeneity on the periods and the damping times of the standing slow-modes in stratified solar coronal loops was studied either (e.g., \citet{ref:Abedini2012}). \citet{ref:Fathalian2019} estimated the loop temperature using the intensity ratios and the AIA response functions in different wavelengths. 
Different emission measure (DEM) computations and methods have been developed to estimate the temperature in the corona, which led to various discussions. \citet{ref:Schmelz2010} analyzed a coronal loop, which was observed on 2010 August 3, by AIA. They took some differential emission measure (DEM) curves, claiming a multithermal rather than an isothermal DEM distribution (for the cross-sectional temperature of the loop). After that, \citet{ref:Aschwanden2011} criticized the method of background subtraction which Schmelz et al. had applied. They claimed that the background subtraction method caused their inferred result of a multithermal loop. \citet{ref:Aschwanden2011} analyzed a set of hundred loops and understood that 66$\%$ of the loops could be fitted with a narrowband single-Gaussian DEM model. In this regard, some attention was paid to the instrumental limitations and ability of AIA and \citet{ref:Guennou2012a,ref:Guennou2012b} discussed on the accuracy of the differential emission measure diagnostics of solar plasmas in respect of the AIA instrument of SDO. The abovementioned controversy of whether the cross-field temperatures of coronal loops are multithermal or isothermal, continued by \citet{ref:Schmelz2013} (similar to \citet{ref:Schmelz2011}). They analyzed twelve loops to understand the cross-field temperature distributions of them and reveal the loops' substructure. Based on their achievements, the warmer loops entail broader DEMs. Thereafter, \citet{ref:Schmelz2014} found indications of a relationship between the DEM weighted-temperature and the cross-field DEM width for coronal loops. They argued that cooler loops tend to have narrower DEM widths. This could imply that fewer strands are seen emitting in the later cooling phase, which they claim could potentially resolve the abovementioned controversy. In this subject, \citet{ref:Aschwanden2015} (as well as 2013 \citep{ref:Aschwanden2013}) developed a method to extract the loop temperature which is based on Gaussian fit for Differential Emission Measure, named spatially-synthesized Gaussian DEM forward-fitting method (DEM hereafter).

This paper aims to analyze and compare thermal oscillations of coronal loops in flaring and non-flaring active regions, 11283 and 12194, respectively. The contents of this paper are as follows: In section \ref{sect:Obs}, data, we introduce the considered flaring and non-flaring active regions and describe the data employed and the time and properties of the flare, occurred in the active region. In section \ref{sec:Analysis}, we explain the method we use to analyze the time-series of temperatures in different strips of the loops. Section \ref{sec:R} is specified to our results, obtained related to flaring and non-flaring regions. In section \ref{sec:S} we briefly state a summary of this work.

\section{Data}\label{sect:Obs}

We investigate the thermal structure and treatment of loops in a flaring region to see if it follows the transverse oscillations of the loops, and we examine the thermal fluctuations at the flare time. For this purpose, we select a high energy flare x2.1 which the transverse oscillations of two loops of it have been analyzed by \citet{ref:Jain2015}. They analyzed intensity variations in the wavelength 171 $\AA$ in two coronal loops of this region and detected obvious transverse oscillation with periods of roughly 2 minutes and decay times of 5 minutes for these loops at the flare time. To see the specific thermal properties of the flaring loops, as a blind test, we select a non-flaring active region, extract its loops and analyze their thermal treatment. Then we compare the temperature treatment of the loops at the flaring region with the loops of the non-flaring region to see the differences. 

The temperature analysis done here uses EUV images from the AIA onboard the SDO. AIA has ten different wavelength channels, three in white light and UV, and the other seven in EUV channels. Between these seven, the 304 filter, which is mostly sensitive to chromospheric temperatures (in order of $T=10^{4.7}$K), not the corona, is not taken into account (Aschwanden et al. 2015). Therefore, we consider the images of the events in the six wavelengths (94, 131, 171, 193, 211, 335 $\AA$). These are covering the coronal temperature range from $T\approx 0.6$ to T $\geq 16MK$.

The two below data sets are finally selected to study thermal variations and coronal loops oscillations in flaring or non-flaring active regions. A few distinct loops are visible in the regions. Finally, these loops are chosen: 
\begin{itemize}
\item[--]
Three loops of the x-flaring active region 11283: Observationally, the X-class flares are rarely happening around the loops with the specification we are looking for. So this selected LOS X-flare, which occurs near the loops is of rare cases. We consider EUV images of NOAA AR 11283, in the time period of 22:10UT till 23:00UT of 2011 September 6 with the cadence of 12 sec. This period of time is selected since no other flare is happening during it. A few distinct loops are visible and follow-able here during this period. Loop shapes in our active region change permanently; therefore, it is difficult or impossible to follow a loop over a very long time. Hence, it is not useful to extend the time interval of this region to the time before the flare. The transverse oscillations of two loops in this region were analyzed before by  \citet{ref:Jain2015}. We mark these loops by A and B in Figure~\ref{Fig1} b. They detected fundamental mode oscillation with periods of roughly 2 minutes and decay time of 5 minutes for these loops. We are curious to see the loops' thermal oscillations (if any) or thermal fluctuations in this condition. Figure~\ref{Fig1}a (left) displays AR 11283 and the area, indicated by the white box is featured in a zoom-in view in Figure \ref{Fig1}.b (right)  and the five selected parts of the center of the three chosen loops are shown by red lines (the movie of the region is available in this link). As it is clear in the movie, these three loops oscillate together and their oscillations decay simultaneously. The center of figure \ref{Fig1}.a is coordinated at (230, 165) arcsec and its width and height are $450^{''}\times 456^{''}$ ${/}750\times 775$ pixels. The flare occurring in this active region is an X2.1 class flare located close to the disk center at latitude 14$^\circ$ north and longitude 18$^\circ$ west (269.9 arcsec, 129.9 arcsec). This flare initiates at 22:12UT, ends about 22:24UT with the peak at 22:20UT, and associates with a coronal mass ejection (CME) which occurs from 2011 September 6, 21:36:05T to 2011 September 7, 02:24:05T, with the radial velocity of 469 km/s,angular width of 252 deg, and position angle of 275 deg (for more details look at LASCO CME catalogue.) 
\footnote{Based on data on these WebSites: https://solarflare.njit.edu/webapp.html, and https://www.swpc.noaa.gov/} 
%\add{We can be sure that the source of this CME is AR 11283.} (\citet{ref:Romano2015})

\item[--]
Three loops of non-flaring active region 12194: As a blind test, we select three loops of the non-flaring (nonf hereafter) active region 12194 in the smooth time period of 08:00:00UT till 09:00:00UT of 2014 October 26. The center of figure \ref{fig2}.a is coordinated at (0, -264)  arcsec and its width and height are 
$615^{''}\times 615^{''}$ ${/}1025\times 1025$ pixels. We consider the images of the selected area with the cadence of 12 sec in the same six wavelengths mentioned above. These loops are relatively motionless and do not show any transversal oscillation (see the region's movie in the link). We select the loops in such a way that they do not have any crossing over the neighbor loops (in our perspective) during this time. In figure \ref{fig2} the selected loops are distinguished in red in the mentioned active region. The size of the final cut of non-flaring region (represented in the right) is $351\times 401$ pixels. 
\end{itemize}
The data set are primarily downloaded at level 1 with a pixel resolution of $0.6$ arcsec. 
We use the standard $aia{\_}prep.pro$ subroutine available in SDO package SolarSoftWare library to adjust the screen scale between the four arms of the AIA. This pre-processing step increases the data level from 1 to $1.5$, so that finally no jump or sudden movement is observed in the image series. We also used $drot{\_}map.pro$ subroutine to correct the differential rotation effect. 
 According to the movie made by pre-processed images, the most obvious loops (marked in the abovementioned figures) are selected in each region (with obvious transversal oscillations in the case of the flaring active region). 

\section{Temperature Analysis Method}
\label{sec:Analysis}

We extract the selected loop segment pixels, for each loop, and calculate the normal vectors to each point of the loop's direction. Then by using these data, we straighten each loop in a considered box with the thickness of 15 to 40 pixels (macro-pixels, depending on the available empty area around each loop and the distance to the neighbor loop). The area around the loop is needed for calculations of background subtraction. The selected loop segment is cut in all wavelengths and at the same considered box from the images set. These loop images are necessary entrances for our thermal analysis process. Then the loop is divided into different strips and its best division in terms of pixel intervals is considered. To do thermal analysis, we use the spatially-synthesized Gaussian DEM forward-fitting method founded by \citet{ref:Aschwanden2015}.

The images in the above six wavelength filters are considered to calculate the temperature in each strip of the loop. The DEM function is considered a single-Gaussian function relative to the temperature determined by the forward fitting method. To obtain the temperature for each loop, we divided the loop into narrow strips, and then the intensity flux was averaged over each strip. The number of each strip is displayed with the index i. One of the usual methods to subtract the background from observed data is fitting a single-Gaussian cospatial function with a linear function on the flux profile. The DEM for each strip is considered to be single-Gaussian DEM in terms of the logarithm of the temperature, which has three free parameters \citep{ref:Aschwanden2011}:

\begin{eqnarray}
DEM_{i}=\frac{dEM_{i}}{dT}=EM_{p,i}\exp{(-\frac{[\log{(T)}-\log{(T_{p,i})}}{2\sigma^{2}_{T,i}})}.
\end{eqnarray}	
In which, $T_{p,i}$ is the DEM peak temperature, $EM_{p,i}$ is the peak EM function, and $\sigma_{T,i}$ 
is the logarithmic width of the temperature for that strip. To calculate the background-subtracted fluxes (for each strip) we use Eq.6 of \citet{ref:Aschwanden2011} (in below):
\begin{eqnarray}
F_{0\lambda}=\int{\frac{dEM(T)}{dT}R_{\lambda}(T)dT}=\sum_{k}EM(T_{k})R_{\lambda}(T_{k}).
\end{eqnarray}	
Here, $R_{\lambda}(T)$ is the instrumental temperature response function of each wavelength filter $\lambda$, which is obtained by the code $aia{\_}get{\_}response.pro$ in the SSW package. As time has passed, the AIA response functions calibration has partly changed. Here, we use the updated calibration of the temperature response functions, for each of the AIA temperature filters, according to the CHIANTI Version 2019 code available in the Solar SoftWare (SSW). After forward-fitting the Gaussian DEM to the background-subtracted observed fluxes in multiple wavelengths, the three-fitting parameters, temperature width ($\sigma_{T,i}$), peak of temperature ($T_{p,i}$), and peak emission measure ($EM_{p,i}$) are found by minimizing $\chi^2_{i}$. 

Our data sample is uneven because of omitting some damaged images in between. Therefore to analyze the temperature oscillations, we use the Lomb-Scargle method. This method is developed to use the technique periodogram, in the case where the observation times are unevenly spaced \citep{ref:Scargle1982}. The Lomb-Scargle periodogram method is useful in cases where the periodicity of data treatment is not immediately apparent. This method allows efficient computation of a Fourier-like power spectrum estimator from unevenly-sampled data, resulting in an intuitive means of determining the period of oscillation \citep{ref:VanderPlas2018}. Therefore we use Lomb-Scargle Periodogram to evaluate and estimate the efficient periods of temperature oscillations in our loops. We select the first period related to the highest power frequency, which is obtained by this method.We considered the achieved periods with the highest significances and amplitudes. The most significant (highest) periods observed in temperature (minute) for flaring and non-flaring loops are listed in Tables 1 and 2, respectively. To estimate the significance of the periods, we computed the probability values (p-values). In the Lomb-Scargle method, the significance returned here is the false alarm probability of the null hypothesis, i.e., as the data is composed of independent Gaussian random variables. Accordingly, low probability values (p-value less than 0.05) indicate a high degree of significance in the associated periodic signal. 

\section{Results}
\label{sec:R}
\subsection{Temperature Analysis of Flaring Active Region Loops}

Thenceforth the temperature time-series of different strips of the selected loops are calculated using the method described in section 3. In the following figures, the vertical axis shows the logarithm of the temperature and the horizontal axis shows the time duration. To be comparable by eyes, all the forthcoming figures (which show the loops temperature oscillations) have been co-scaled in the ($\log$) temperature range of $5.7$ to $6.9$. The color maps are shown for each temperature map. 
Loops A, B1, B2, C1, and C2 are subdivided into 25, 11, 8, 12, and 6 strips, respectively. Each strip's length is equal to 4 pixels (macro-pixel), for all loops in this paper. For brevity, a few strips' temperature oscillations are presented here. Figure \ref{fig3} displays the time-series of temperature oscillations for the first 3 strips of Loop A, and first 2 strips of loops B1. We calculated the errors for each point (temperature) but removed in the presentation to avoid overcrowding of the figures. As we observe in Figures \ref{fig3} and \ref{fig4}), the temperature oscillations are started and increase around 22:12 before the flare peak time (22:20) and are mostly continuing after the flare ended (22:24). These temperature oscillations follow the transverse loop oscillations observed by \citet{ref:Jain2015}. As Jain et al. reported, LoopA and B have a transverse oscillation with periods of roughly 2 minutes and decay times of 5 minutes, starting at 22:18 around the flare peak time (23:20) and decaying after the flare ended (22:24). So as we observe, the temperature oscillations in these flaring loops happen before the start of their transverse oscillations and are continuing even in the time interval after the transverse oscillations decay. Although the temperature oscillations do not decay as rapid as the transverse oscillations do, and conversely, the loop temperature increases at the end of the oscillating mode (see Fig.\ref{fig4}, the temperature map of the loop A, for instance) 

We calculate the temperature oscillations periods, using Lomb-Scargle method. We consider the thermal oscillations periods with the highest significances. As this method shows, the most powerful period in the range of data time-series (listed in Table\ref{table1}) are from 7 to $28.4$ minutes observed in the strips of the marked loops of this flaring region. These loops of flaring region also show some short periods in temperature oscillations which some are less than 10 minutes (listed in Table\ref{table1}). These short periods are more frequently observed in the loops of the flaring active region. Such short periods are very scarce for the loops of the non-flaring active region (compare Tables\ref{table1} and \ref{table2}). 

The first column in Table\ref{table1} is the number of every strip along the loop. The second column is the period of the most powerful frequency observed for the loop strips, calculated by the Lomb-Scargle method. The third column shows the maximum of $\log(T)$ minus its minimum in each strip. The columns of Table\ref{table2} are exactly the same as Table\ref{table1}; the only difference is that Table\ref{table2} is for the non-flaring loops.

The loop A, has the length of $42.3$ (Mm) which is the length of the selected part of the loop marked in Figure \ref{Fig1}.b. The mean of the parameter (Max($\log{T}$)-Min($\log{T}$)) for the strips of loop A is $1.21$. Mean of the temperature ($\log$) of this loop over time is $6.15 \pm 0.25$. The loop B1, divided into 11 strips, has the length of $20.24$ (Mm). The mean of (Max($\log{T}$)-Min($\log{T}$)) and the mean of the temperature for this loop are, $1.10$, and $6.28 \pm 0.22$ respectively. The loop B2, which has 8 strips, with the length of $15.61$ (Mm), has the mean temperature ($\log$) of $6.21 \pm 0.21$. The mean of (Max($\log{T}$)-Min($\log{T}$)) is $0.81$ through this loop segment. The loops C1 and C2, divided into 12, and 6 strips, have the lengths of $22.08$ and $11.06$ (Mm), the mean temperatures of $6.25 \pm 0.22$, and $6.14 \pm 0.25$ (log), and the mean (Max($\log{T}$)-Min($\log{T}$)) of $1.48$, $0.88$, respectively.

We observe that despite the temperature oscillations, the flaring loops show a temperature rise at the end of the considered time interval (figure\ref{fig3}). As their temperature maps also show, the oscillations follow with a relatively sensible rise in the final temperature of the loop segments (Figures \ref{fig4}). Although in the case of the transverse oscillations, the loops oscillate as the flare occurs and then the oscillations decay and stop, in the case of temperature oscillations, the temperatures of the various strips of the loops oscillate and at the end of the flare occurrence, they get to a relatively higher value of temperature in average.

Figure \ref{fig4} shows the temperature maps of the flaring loops A, B1, B2, C1, and C2, respectively as a time series. In each plot, the vertical axis is the distance along the loop segment in Mm, and the horizontal axis shows time. The color bar (in the left) shows the temperature range. Each separated grid part on the map is standing for one strip. Figure \ref{fig4} shows that the temperature for most of the strips increased, bypassing a few oscillations. Before the end of the time duration, some strips become hotter (yellow ones) and some cooler (blue ones). The loop B1 is colder at the early times of the duration and becomes hotter at the middle and end times with a swing to lower temperatures again (see Fig. \ref{fig4}). There are some temperature fluctuations at the middle times (the red and green stripes) while at the end the strips temperatures are smoother with less fluctuations. The temperature map of the loop segment B2 (Fig.\ref{fig4}) shows that at the beginning of the time duration, the first strips of the loop are hotter, and the last ones are colder, but at the end times this pattern is reversed in this loop segment. In loop segment C1 (Fig.\ref{fig4}), the temperature fluctuations are mainly observed to start after the end of the flare (22:24), and at the end time (23:00) the temperature is much higher than the beginning. The temperature is increasing after the flare time (22:24) for the loop C2 either (see Fig.\ref{fig4}). This happens with some oscillations in the strips' temperatures. So as figure \ref{fig4} shows, the temperature increases with some fluctuation in most of the flaring loops' strips after the flare time. According to these temperature maps, the temperature fluctuations in the flaring loops are increasing at the flaring time and around 20 minutes after that. 

We expect the flaring loops to cool down as a result of heat conduction and radiative cooling. Hence this relative temperature increase should be scrutinized. 
As we probed, this temperature rise is also followed in intensity time-series. As the intensity time-series show, the related intensity in the Loop A of the flaring AR increases at the end of the time duration. To be assured, the authors also checked the wavelength of Fe   $\footnotesize{XVIII}$ which has a peak formation temperature of $7\times 10^{6}$ $^{\circ}K$ (\citet{ref:Ugarte2014}). 
By using the method developed by \citet{ref:Warren2012} the contribution of the Fe $\footnotesize{XVIII}$ emission line can be isolated from the AIA 94 $\AA$, to analyze the evolution of hot plasma in the loops. We do it to omit the contamination from the cooler plasma (mostly around 1MK) which also contributes to this AIA channel \citet{ref:Boerner2012}. 
This is done by subtracting the contaminating warm (i.e., around 1MK) component to the bandpass. This warm contribution is calculated from a weighted combination of the emission from the AIA 171 $\AA$ and 193 $\AA$ channels dominated by Fe $\footnotesize{X}$ and Fe $\footnotesize{XII}$ emission, respectively. 
This intensity analysis is done directly and it has not gone through any other process like the thermal analysis. For this purpose, we applied the formulation (1) used by \citet{ref:Li2015}. Plots in Figure \ref{figIntF} show the intensity map, and the mean intensity variation of the wavelength Fe $\footnotesize{XVIII}$, for Loop A of the flaring region, respectively. As these plots show, this intensity is also higher at the end of the time duration in respect of the flare time. 
It seems to us that the expected cooling has not occurred in these flaring loops yet, even after the flare occurrence in the probed duration due to some plausible reasons. 
We consider that the mentioned simultaneous CME (see section\ref{sect:Obs}) which this flare is associated with could cause this increase in temperature. We can be sure that the source of this CME is AR 11283 (\citet{ref:Romano2015}). This CME is in our flare region, hence the loops receive energy even after the flare occurrence and it is probably the reason why the expected cooling does not occur.

The thermal oscillations periods obtained the Lomb-Scargle method, do not have the same significance in all strips of the loops, but for most strips of the flaring loops, the significances are very near to one. To be assured about these oscillations, we probed the intensity time-series for each strip of the loops and we observed that this loop's intensities shows intensity oscillations too (i.e., alongside the loop). The most probable dominant periods observed in intensity, for wavelength of 171 $\AA$ is 18.22, and 16.7 min for strips of F-Loop A, 16.7, and 18.22 min for strips of F-Loop B1, 16.70, and 12.52 for F-Loop B2, and 16.7 for F-Loop C1 and F-Loop C2. These periods are in the same order of the observed thermal oscillation periods. The intensity in this time series has not passed any thermal process but still shows oscillation periods close to thermal ones. So we think these results confirm the observation of thermal oscillations.

\subsection{Temperature Analysis of non-Flaring Active Region Loops}
The temperature time-series for different strips of the selected loops of the non-flaring active region 12194 are calculated using the Lomb-Scargle method. In the following figures (Fig. \ref{fig5}), the vertical axis shows the logarithm of the temperature and the horizontal axis shows the time duration. Figure \ref{fig5} displays the time-series of temperature variations for the first two strips of the non-flaring Loops A, and B. These figures are all co-scaled in the range of $5.7$ to $6.9$ for the logarithm of temperature (like the flaring loops range). The most powerful periods, observed in most of these non-flaring loops' strips (listed in Table\ref{table2}) are from $8.5$ min. to 30 min. Comparing the periods of the loops in the flaring region (Table\ref{table1}) with the non-flaring one (Table\ref{table2}), we see that the temperature periods of the flaring loops have lower values on average and have more diversity than the non-flaring ones. As Tables \ref{table1} and \ref{table2} show, the mean temperatures of nonf-loops are lower in comparison with the f-loops, a fact we also expected from common sense. The parameter (Max($\log{T}$)-Min($\log{T}$))  in nonf-loops' strips is less than that for the flaring loops' strips.

Nonf-loop A, divided into 11 strips, has the length of $19.91$ (Mm) which is the length of the selected part of the loop marked in Figure \ref{fig2}b. The mean of (Max($\log{T}$)-Min($\log{T}$)) for the strips of nonf-loop A is $0.81$. Mean of the temperature ($\log$) of this loop segment over time is $5.93 \pm 0.10$. Nonf-Loop B, divided into 6 strips, has the length of $11.11$ (Mm), and the mean temperature ($\log$), and the mean of  (Max($\log{T}$)-Min($\log{T}$)) for this loop are, $5.99 \pm 0.13$ and $0.62$ respectively. Nonf-loop C, which has 5 strips, with the length of $10.13$ (Mm), has the mean temperature ($\log$) of $5.82 \pm 0.12$, and the mean (Max($\log{T}$)-Min($\log{T}$)) of $0.56$. 

The first highest period observed for the temperature oscillations of these non-flaring loops' strips is reported in Table\ref{table2}. As we observe the temperature periods in these non-flaring loops are mostly longer than those of the flaring loops (compare the values listed in Table\ref{table1} and Table\ref{table2}). Therefore the temperature oscillations of these loops are a little slower than the flaring ones. 

Figure \ref{fig6} shows the temperature maps of the non-flaring loops A, B, and C, respectively as a time series. In each plot, the vertical axis is the distance along the loop in Mm, and the horizontal axis is the time. The color bar in the left shows the colors considered for the temperature range. Each separated colored part in the map is one strip. These color maps are plotted totally at the same color range of the loops of the flaring region either. 

As figure \ref{fig6} shows, the strips' temperature of these non-flaring loops have fewer temperature fluctuations and are smoother in comparison with the flaring ones (Fig. \ref{fig4}). Furthermore, that much increase in the temperatures of the strips, which was obvious in the loops of the flaring region toward the end times, is not observed here. The temperatures are also totally lower in the nonf-loops in comparison with the flaring loops. Conversely, it seems that different strips of the non-flaring loops have relatively more similar temperature fluctuations.

As figure \ref{fig7} shows, the peaks of the observed temperature periods for the loops' strips of the flaring active region (blue ones), and non-flaring active region (red ones), are around 18 minutes, and 30 minutes, respectively. The temperature periods' diversity is higher in the loops'  strips of the flaring active region, and shorter temperature periods (less than 10 minutes, nearer to the transverse oscillations periods) are observed in the case of the flaring loops' strips in comparison with the non-flaring ones.  And figure \ref{fig8} shows that the increasing and decreasing of temperature range, or the difference between maximum and minimum of the temperature value (max($\log(T)$)-min($\log(T)$)), is much higher on average for the loops' strips of the flaring AR in comparison with the loops' strips of the non-flaring one.  

\section{Summery}   \label{sec:S}

\begin{figure}[htp]
  \centering
 \begin{tabular}{cc}
    \includegraphics[width=50mm]{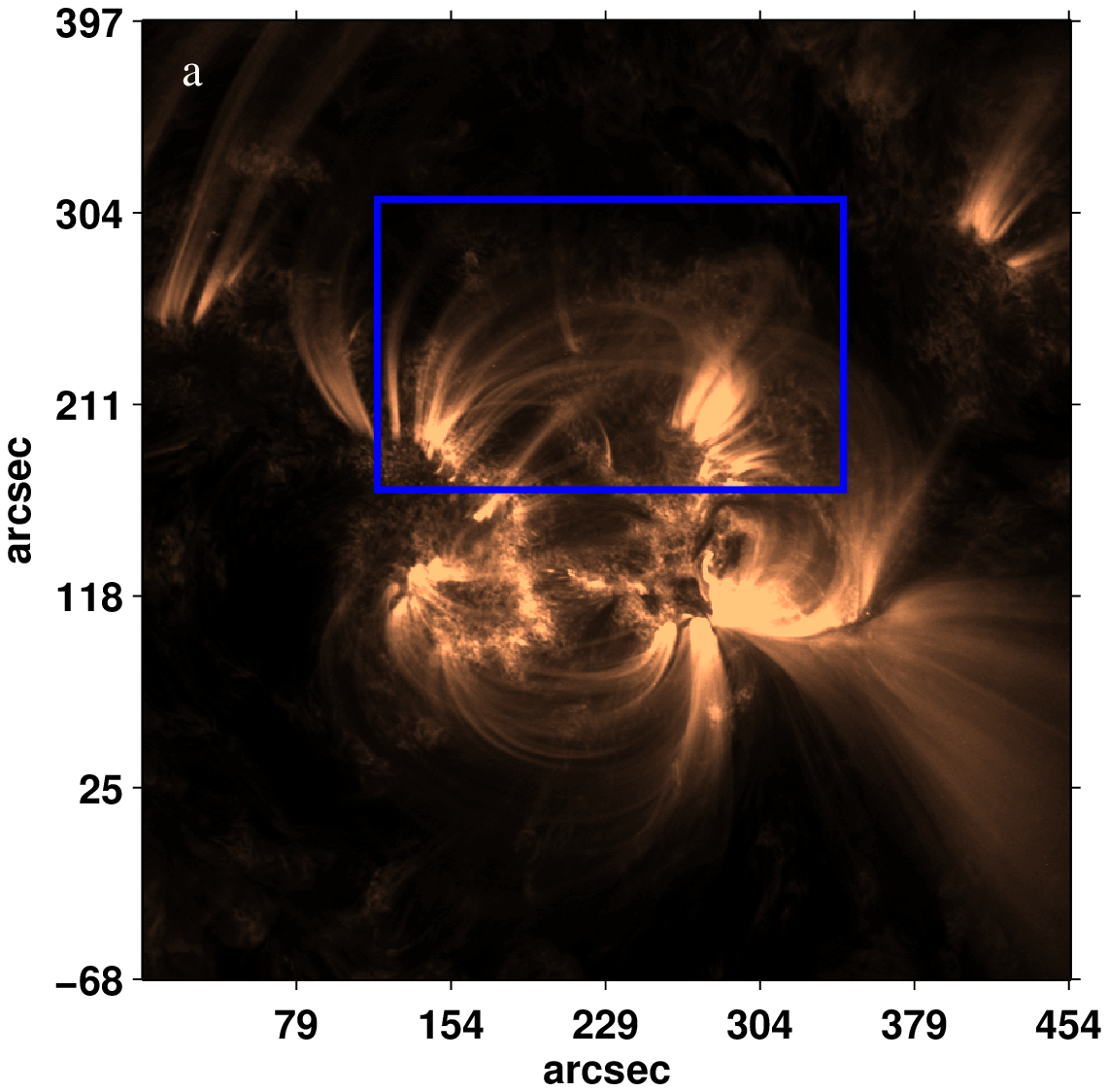}&
    \hspace{1cm}
    \includegraphics[width=40mm]{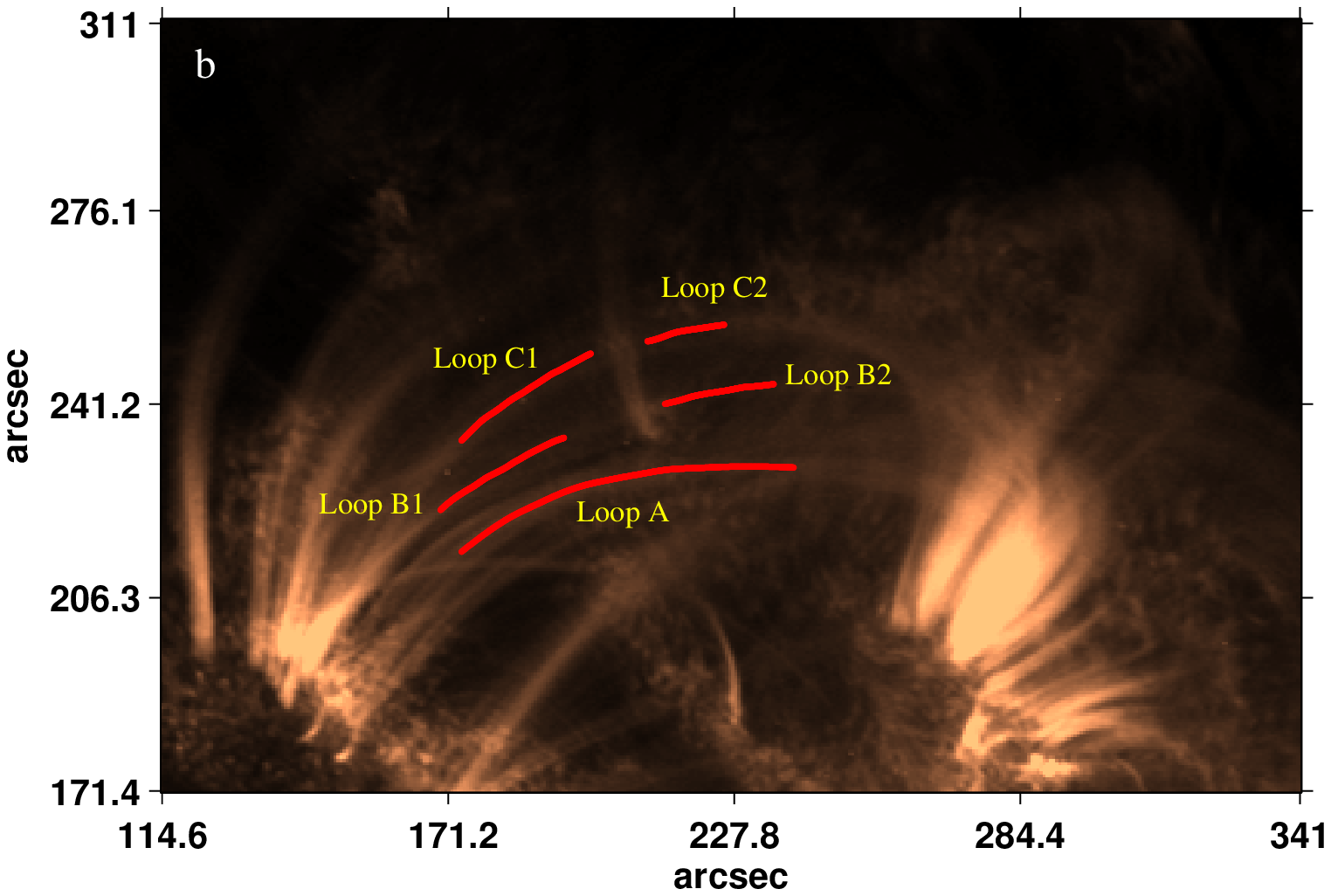}\\
 \end{tabular}
\caption{(a) AIA image of the AR 11283 on 2011 September 6, 22:10 UT as seen in the 171 $\AA$  filter. (b) Zoom-in view of the area marked by a box in the left. The selected loops are distinguished in red. The loops A and B are the same loops studied by \citet{ref:Jain2015} (see Fig.3a in \citet{ref:Jain2015}).}
\label{Fig1}
\end{figure}

We reported the temperature oscillations of coronal loops of a flaring active region. We selected the flaring active region 11283 to investigate the thermal structure and treatment of its loops. This region includes a high energy flare x2.1 and the transverse oscillations of two loops of it have been analyzed before by \citet{ref:Jain2015}. They analyzed intensity variations in the wavelength 171 $\AA$ in two coronal loops of this region and detected obvious transverse oscillation with periods of roughly 2 minutes and decay times of 5 minutes for these loops (loops A and B in Figure.\ref{Fig1}b) at the flare time. We were curious to know if the temperature variations follow the transverse oscillations of the loops, or there is any relation or correlation between them. We also wanted to investigate the thermal fluctuations at the flare time. As a blind test to see the specific thermal properties of the flaring loops, we selected a LOS non-flaring active region (12194), extracted three segments of its loops and analyzed their thermal treatment. Then we compared the temperature treatment of the loops at the flaring region with the loops of the non-flaring region to see the differences. We were eager to observe the probable discrepancies between flaring and non-flaring loops in this respect.

\begin{figure}[htp]
  \centering
    \begin{tabular}{cc}
    \includegraphics[width=50mm]{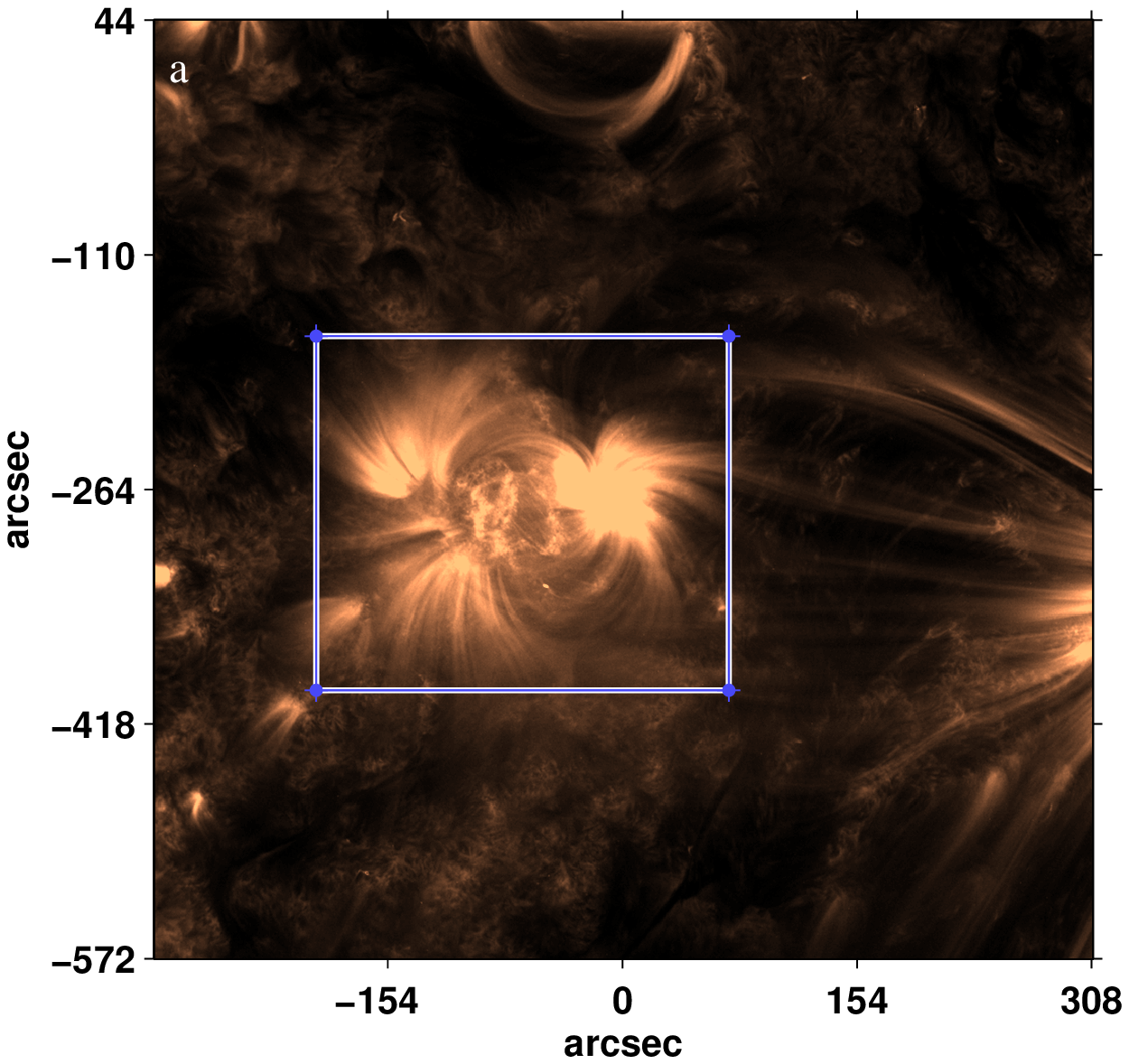}&
    \hspace{1.5cm}
    \includegraphics[width=40mm]{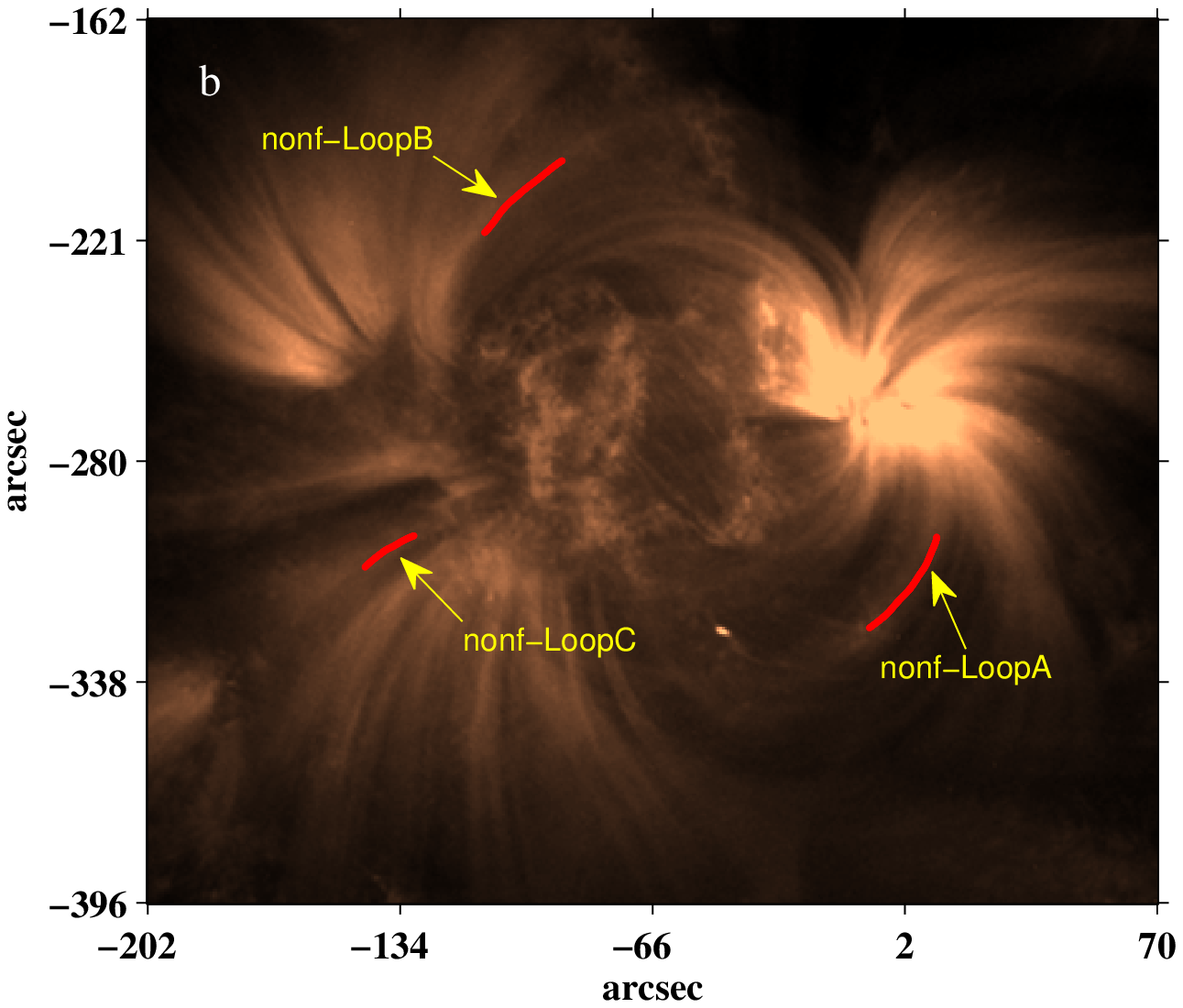}\\
  \end{tabular}
\caption{(a) The NOAA AR12194 on 2014 October 26, at 08:00:00UT in 171 $\AA$ recorded by AIA/SDO. (b) Zoom-in view of the area, marked by a box in the left, the loops are distinguished in red.}
\label{fig2}
\end{figure}

\begin{figure}[htp]
  \centering
    \begin{tabular}{cc}
    \includegraphics[width=120mm, height=60mm, keepaspectratio]{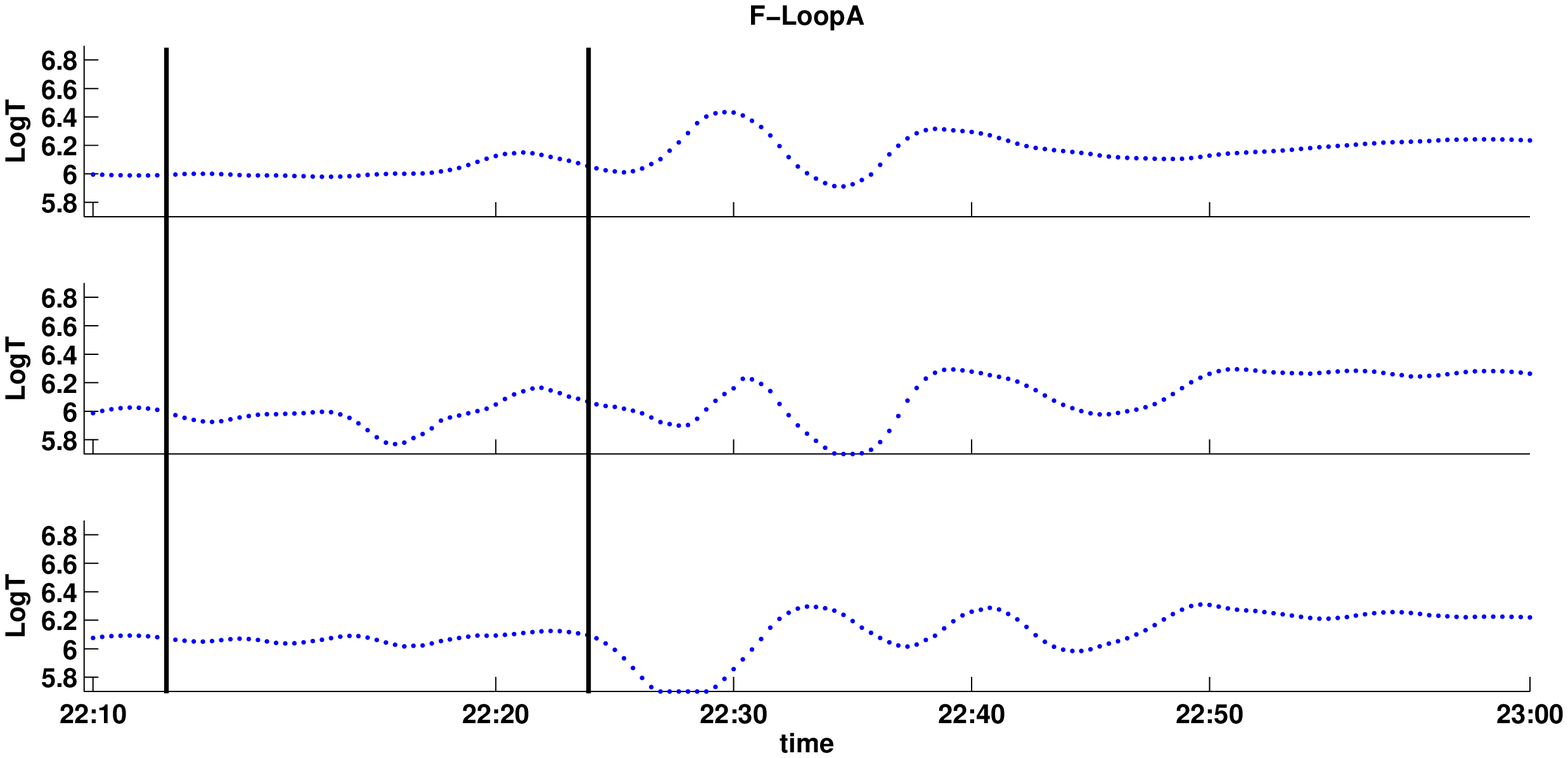}\\
    \includegraphics[width=110mm, height=55mm, keepaspectratio]{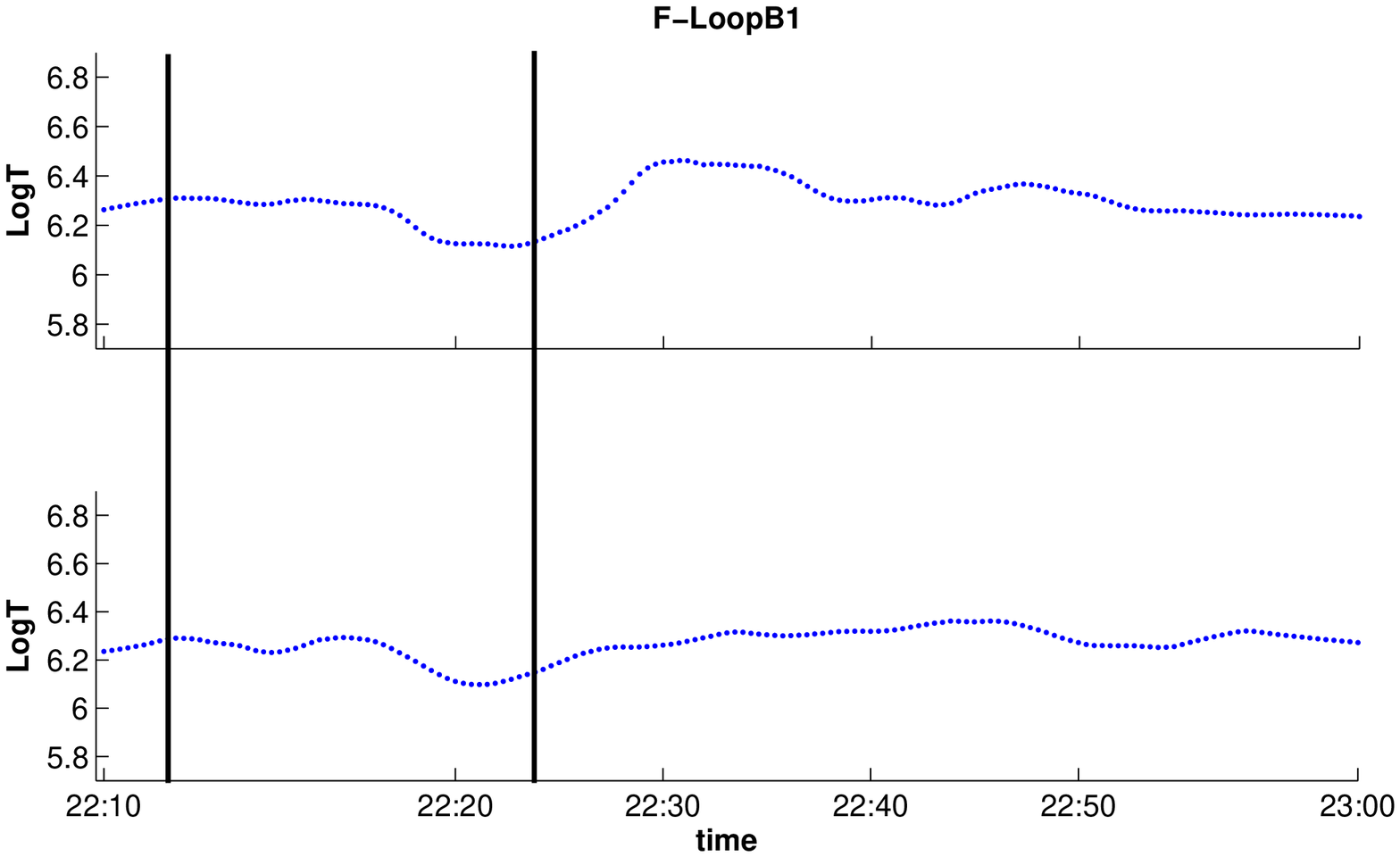}\\  
   \end{tabular}
\caption{From up to down: The time-series of the temperature oscillations for the first 3 strips of Loop A (strip 1 to 3 from top to down), and the first 2 strips of LoopB1. 
 Horizontal axis is the time and the vertical axis is the logarithm of the temperature. The red lines mark the initial and final time of the flare x2.1.}
 \label{fig3}
\end{figure}

\begin{figure}[htp]

  \centering

    \begin{tabular}{cc}
    
    \includegraphics[width=120mm, height=40mm]{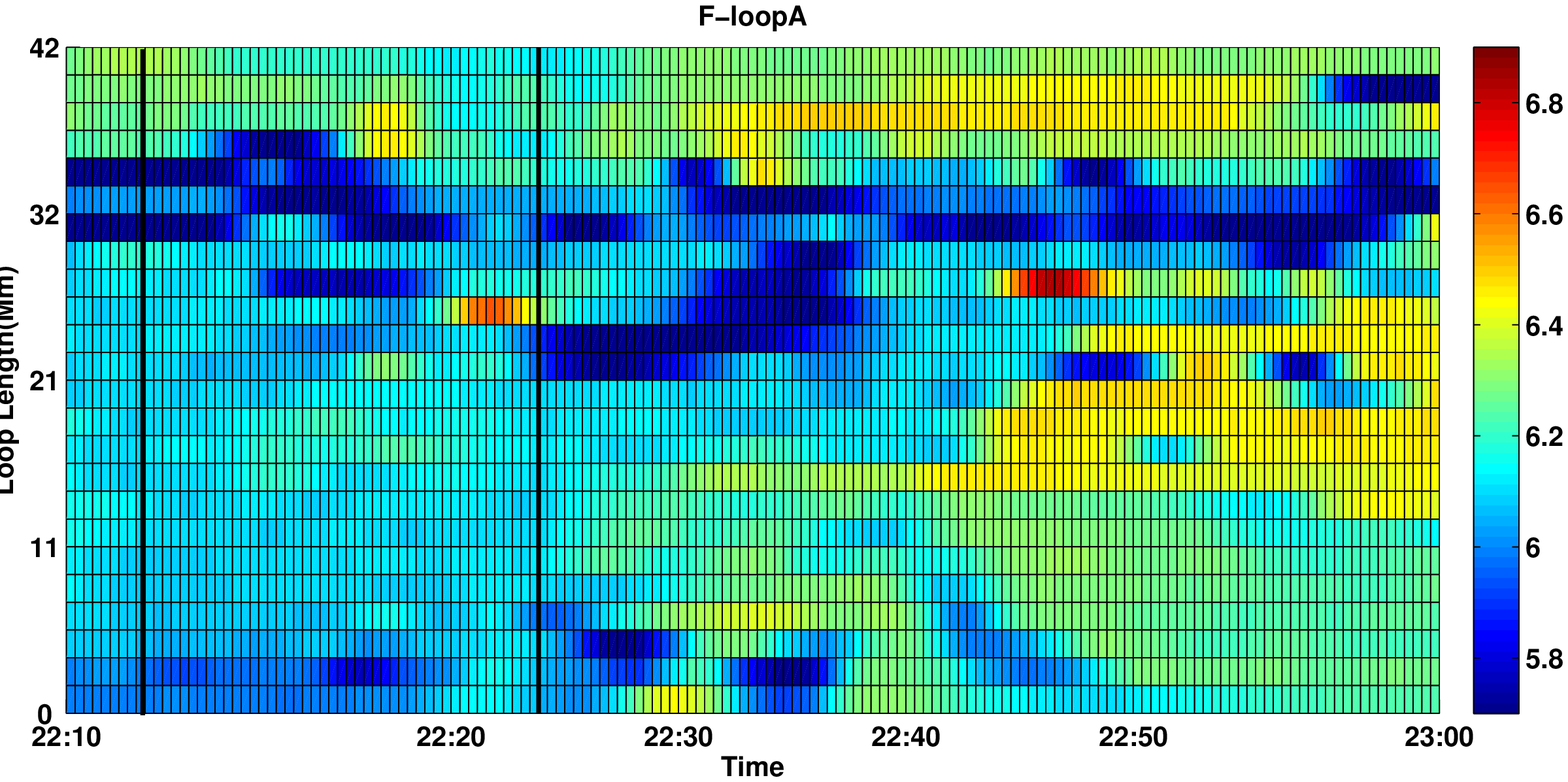}\\
    \includegraphics[width=120mm, height=40mm]{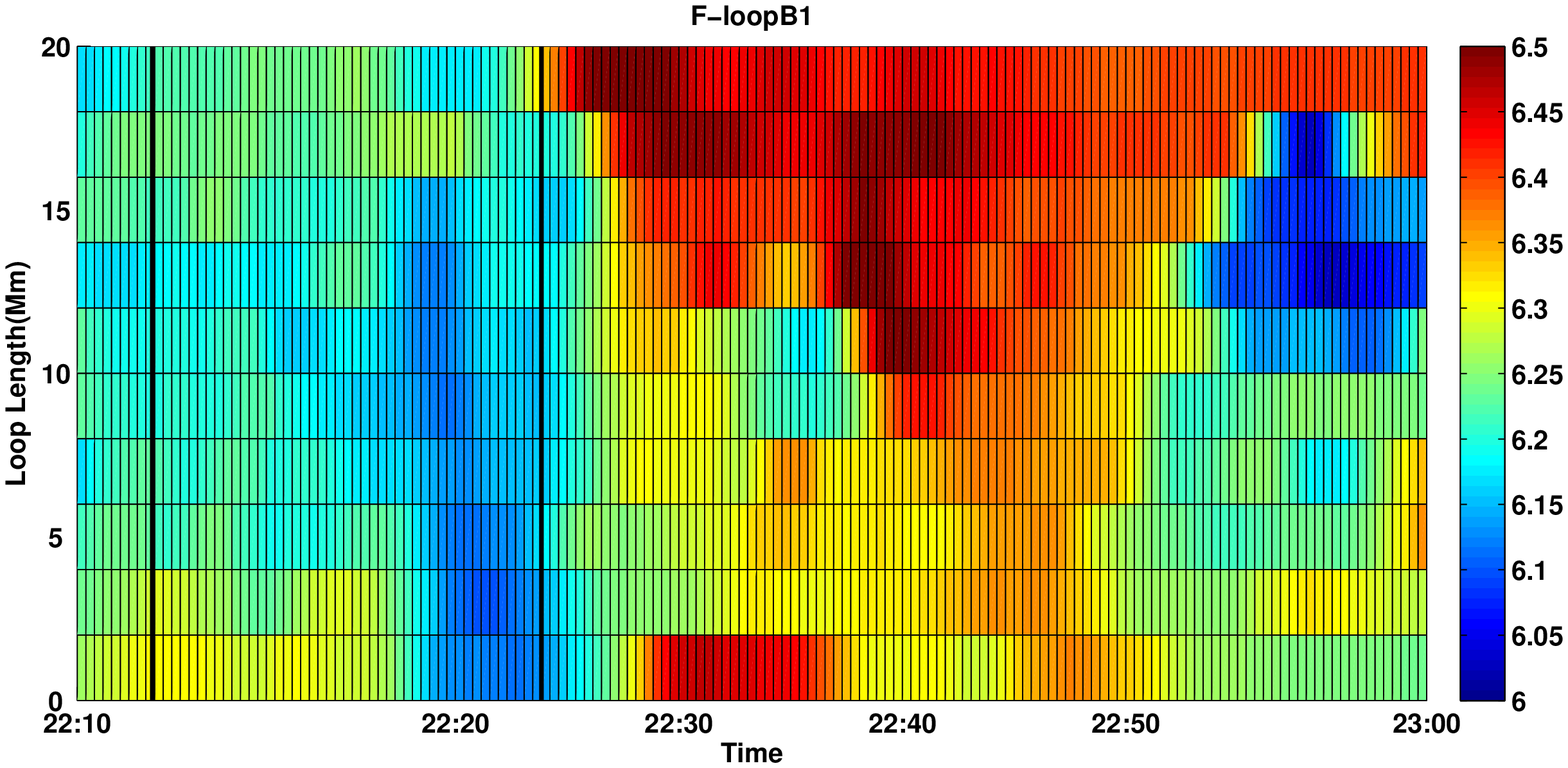}\\
    \includegraphics[width=120mm, height=40mm]{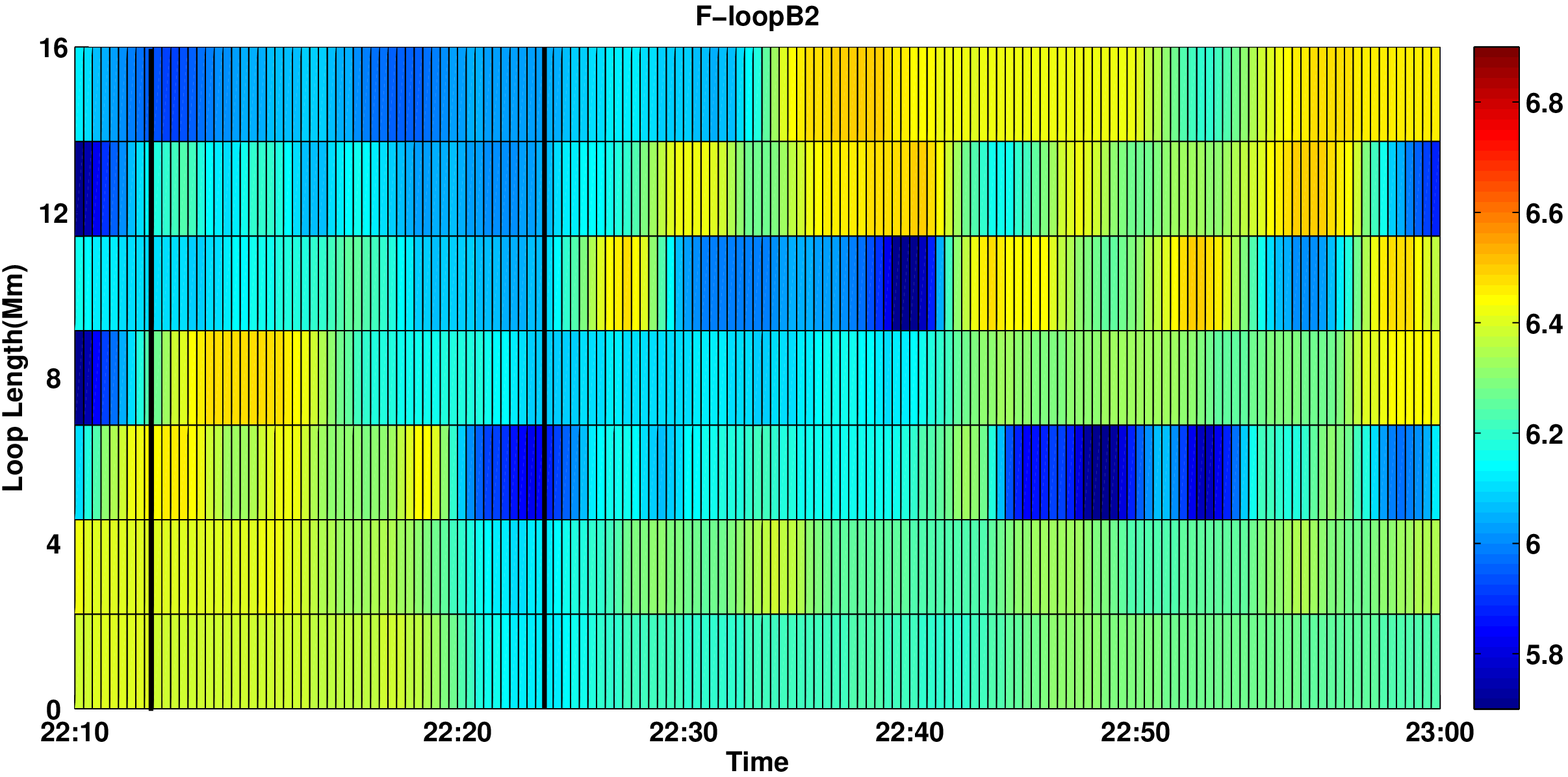}\\
    \includegraphics[width=120mm, height=40mm]{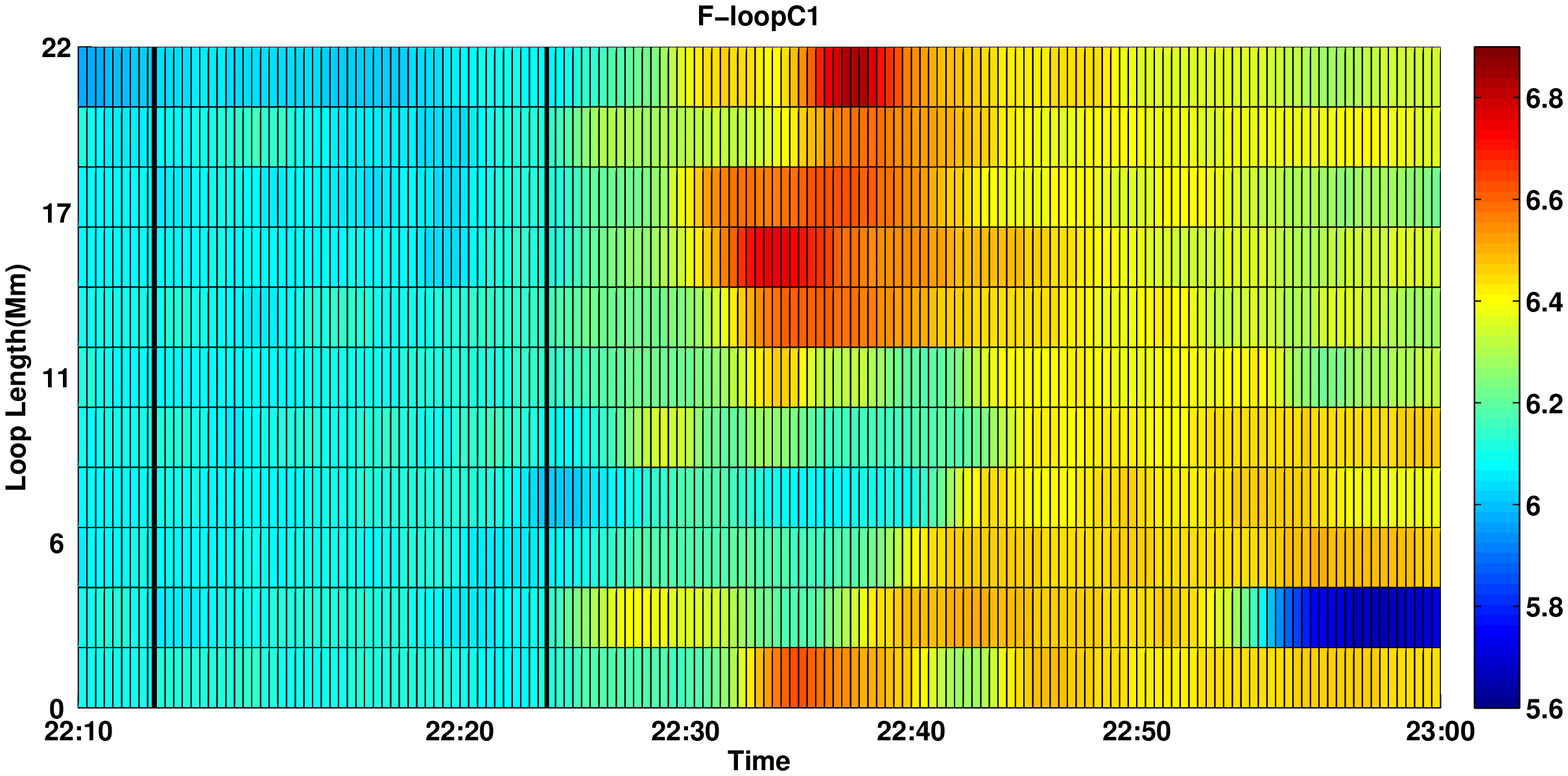}\\
    \includegraphics[width=120mm, height=40mm]{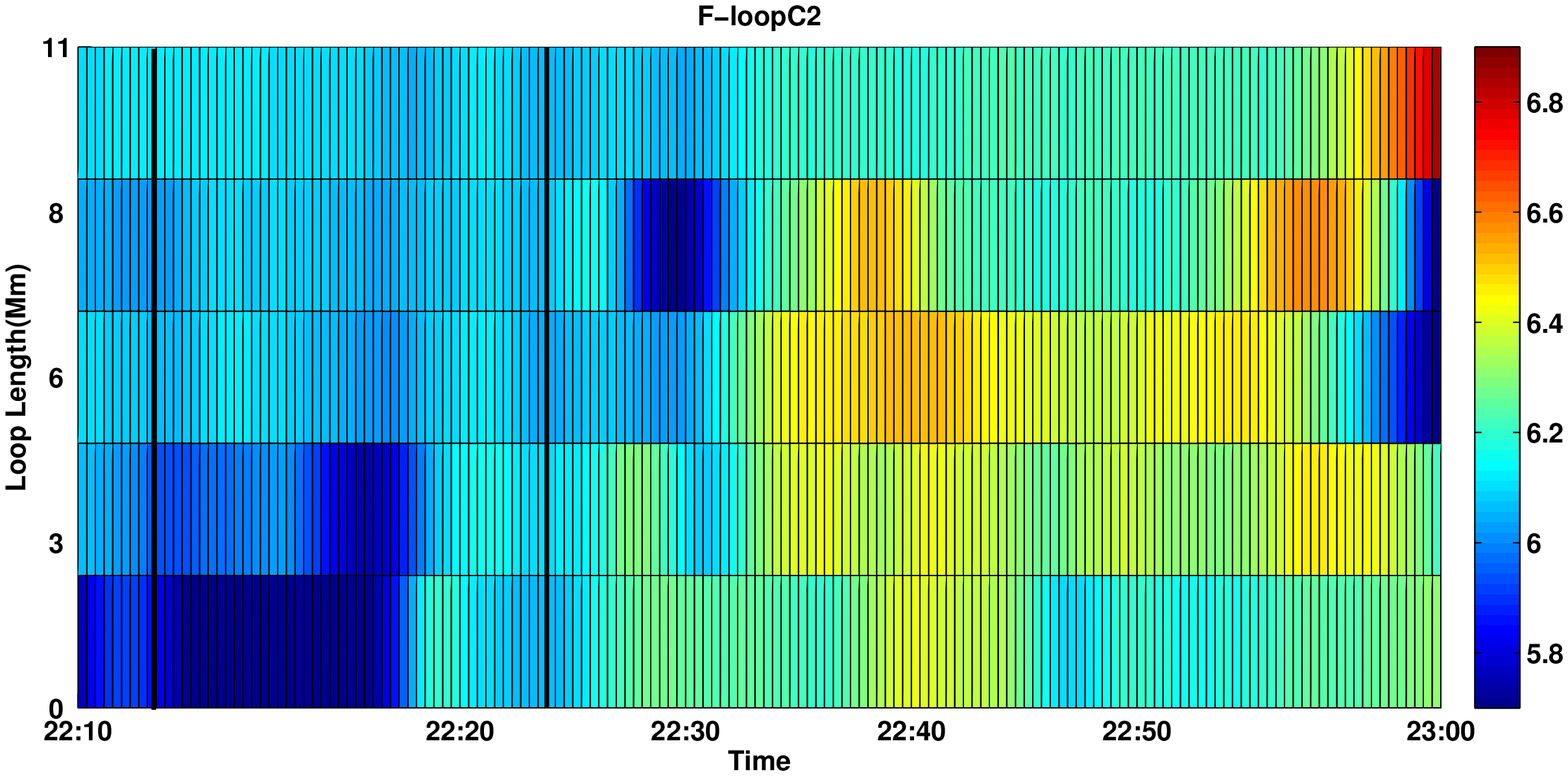}\\
  \end{tabular}
\caption{Temperature map of the flaring loops A, B1, B2, C1, and C2 (from top to down) as a time series. The vertical axis is the distance along the loop in Mm, and the horizontal axis is the time. The colorbar in the left shows the colors considered for the temperature range. }
\label{fig4}
\end{figure}

\begin{figure}[htp]
\centering
\begin{tabular}{cc} 
\includegraphics[width=120mm, height=40mm]{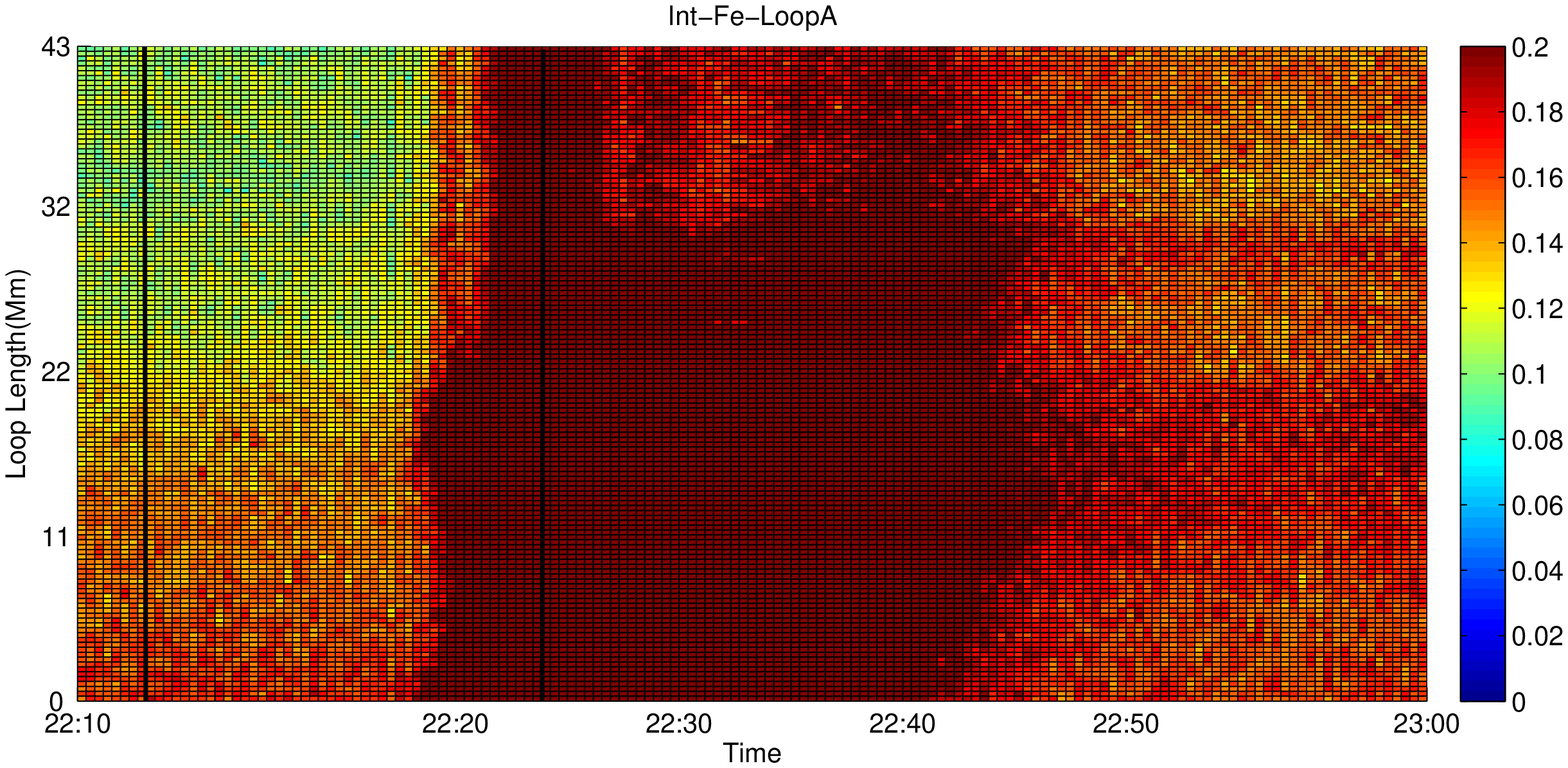}\\
\includegraphics[width=120mm, height=40mm]{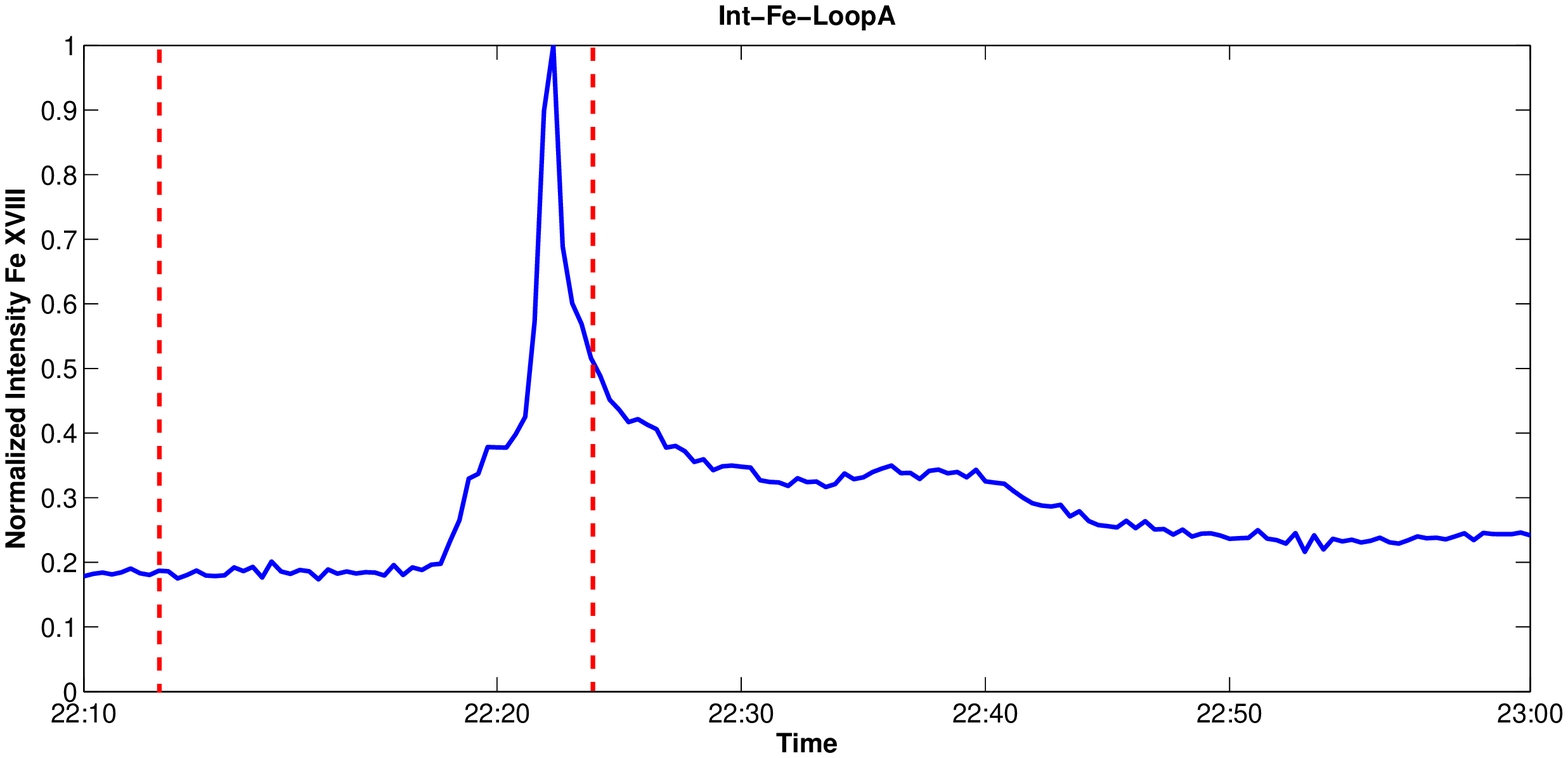}\\
\end{tabular}
\caption{Normalized intensity map of the flaring loop A for the wavelength Fe $\footnotesize{XVIII}$, and mean intensity of Fe $\footnotesize{XVIII}$ (from top to down). The vertical axis is the distance along the loop in Mm for the first plot, and normalized intensity for the second. The horizontal axis is the time. The colorbar in the left shows the colors considered for the Intensity range.
}
\label{figIntF}
\end{figure}

Here we used data of three loops of the flaring active region (AR11283) around the time of the Flare X2.1, from 22:10UT till 23:00UT on 2011 September 6, plus three loops of the non-flaring active region (AR12194), from 08:00:00UT till 09:00:00UT of 2014 October 26 (marked in figures \ref{Fig1} and \ref{fig2}). To calculate the time series of the loop temperature values, we first extracted the loop pixels in each image and then displayed the loop straightly for all the images in the time series of different wavelengths. To do thermal analysis, we used the spatially-synthesized Gaussian DEM forward-fitting method founded by \citet{ref:Aschwanden2015}. We calculated the peak temperatures for each strip of the loops. Then we applied the Lomb-Scargle method to analyze temperature oscillations of the time-series for each strip of the loops.

We observed temperature oscillations which are following the transverse loop oscillations observed by \citet{ref:Jain2015} for the flaring loops. Furthermore, the temperature oscillations in these flaring loops happen before the transverse oscillations start and continue even in the time duration after the transverse oscillations decay. As observed, the temperature oscillations do not decay as rapidly as the transverse oscillations do. Conversely, the strips' temperatures increase at the end of the oscillating mode and a rather sensible rise is observed in the final temperatures of the f-loops' segments. The ranges of the obtained periods are from 7 min. to $28.4$ min. for the flaring loops, and from $8.5$ min. to 30 min. for the non-flaring loops. With the onset of X-flare in the F-loopA, which has a distinct transverse oscillation in the flaring time with period of roughly 2 minutes and decay time of 5 minutes, a temperature oscillation is observed with periods of roughly 10 to $28.5$ minutes in different segments of this loop. And as the transverse oscillation decays in this interval, no special definite decay is observed in its temperature oscillations.

The temperature periods of the flaring loops are rather shorter than the temperature periods of the non-flaring loops. The loops of the flaring region show some short temperature oscillations periods in which some are less than 10 minutes (Table\ref{table1}). These kind of short periods are more frequently observed for the loops of the flaring active region and in the case of the non-flaring ones, are very scarce. We observed that the periods of the flaring loops have more diversity than those of the non-flaring ones. Based on our confined observations,  the non-flaring loops' periods are longer and their temperatures' values are totally lower. So our research showed that thermal structures of the flaring loops differ from the non-flaring ones in the ways described above. As temperature maps show, the temperature fluctuations are increasing at the flaring time and around 20 min. after, in the flaring loops. This happens with some oscillations in strips' temperature. Conversely, it seems that different strips of the non-flaring loops have relatively more similar temperature fluctuations. The temperatures are either higher in average in the flaring loops' segments as expected. The significances of the periods, obtained by the Lomb-Scargle method, are calculated for each strip of each loop and the results show that these significances for the loops' strips of the flaring region are high and close to one, while for the loops' strips of the non-flaring region are less than 0.5. Hence the detected periods in the flaring loops' strips have high significances (near to one) and are oscillations. Whereas the detected periods in the non-flaring loops' strips have less significances in comparison with the flaring ones, and maybe they are just fluctuations.

Using this method for the coronal loops showed that the oscillation modes obtained for the temperatures of the flaring loops are very close to those of the spatial slow-mode oscillations of the coronal loops. 
So the origin of temperature oscillation is probably slow-mode waves. These kind of oscillations often occur in hot coronal loops ($\log(T)>6$) of active regions especially the ones associated with small (or micro-) flares (\citet{ref:Wang2021}). The loops of our flaring active region are also hot loops with the mean temperature above this range. They also show intensity oscillations. Hence we think the above evidence confirms the slow-mode oscillations for flaring loops. The temperature of the non-flaring loops are lower ($\log(T)<6$) and as discussed above, we believe that the observed oscillation-like periods in non-flaring loops should be more probably related to the high amplitude fluctuations.

Comparing the loops of the flaring and non-flaring regions, we observed that the amplitudes of the fluctuations show a discrepancy. Mean of the parameter (Max($\log{T}$)-Min($\log{T}$)) in the FloopA, , FloopB1, FLoopB2, FloopC1, and FloopC2, are $1.21$, $1.10$, $0.81$, $1.48$, and $0.88$, respectively. And for non-flaring region, mean of (Max($\log{T}$)-Min($\log{T}$)), are $0.81$, $0.62$, and $0.56$, for nonfloopA, B, and C respectively. Therefore the values of the quantity mean of (Max($\log{T}$)-Min($\log{T}$)) for these non-flaring loops show a difference from the flaring ones and are lower.

Loops of the non-flaring active region 12194 have a relatively uniform temperature at the beginning of the time interval, which rises slightly at its end. As the Solar Monitor reports in the neighborhood of this region, the flaring active region 12192 exists of which between its multiple flares, there is a $c4.6$ class flare occurring at 9:44UT. Therefore, it could be a possible suggestion that the abovementioned slight temperature rise in the loops of AR 12194 (in the time interval 8:00 to 9:00) originated from the influence of an increase in the energy at the pre-flare conditions exist in the AR 12192.  

Hence as our study shows, the temperature of coronal loops of flaring AR changes in an oscillatory manner. Compared with these non-flaring loops, the flaring loops show higher temperatures on average and higher oscillation periods with higher peaks and deeper valleys. More accurate commentary in this respect requires more extensive statistical research and broader observations.

%%%%non flaring figs
\begin{figure}[htp]

  \centering

    \begin{tabular}{cc}
    
    \includegraphics[width=120mm, height=60mm, keepaspectratio]{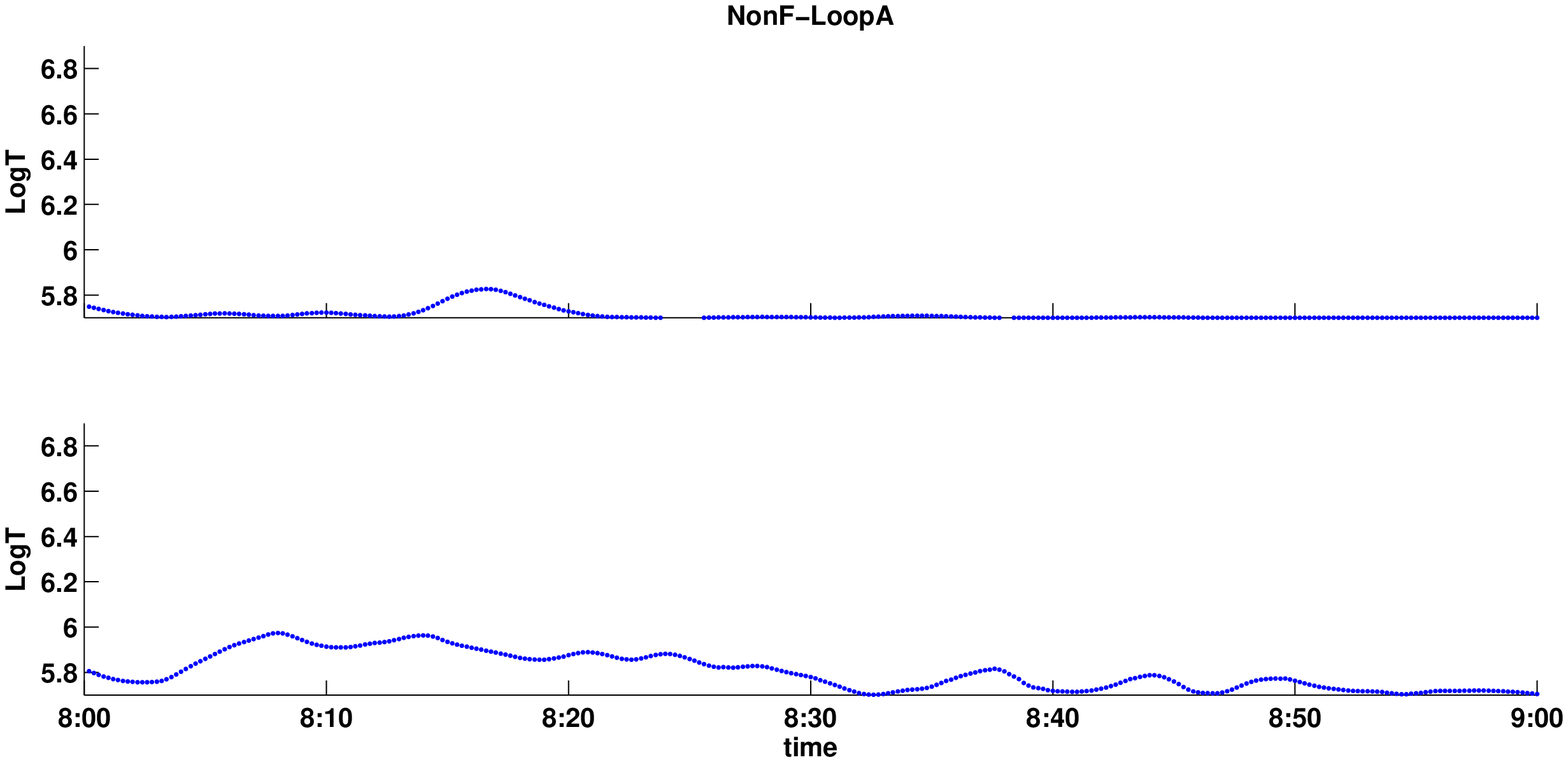}\\
    \includegraphics[width=120mm, height=60mm, keepaspectratio]{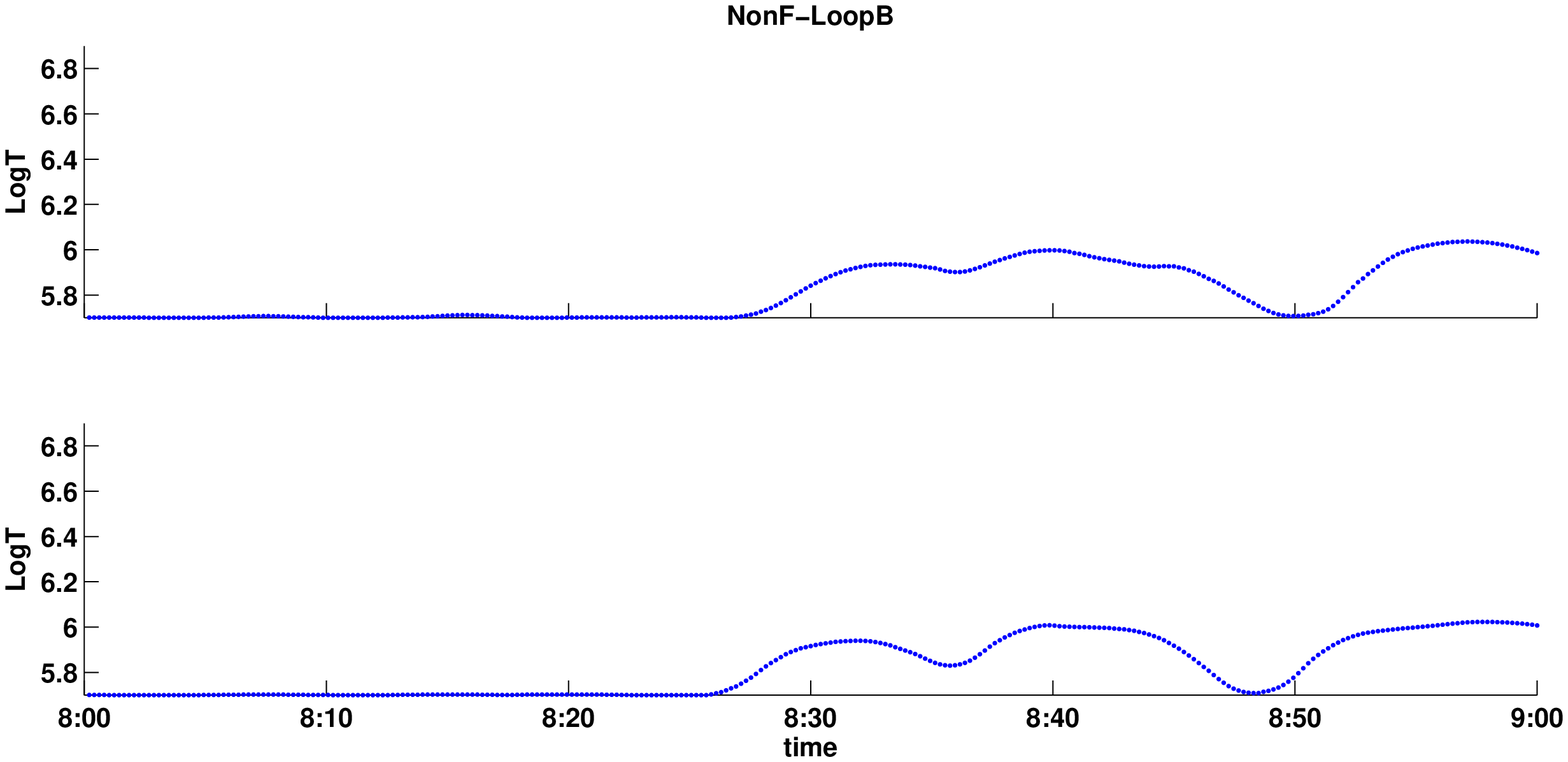}\\
  \end{tabular}
\caption{from top to down: The time-series of the temperature for the first 2 strips (from top to down) of the non-flaring Loops A and B. Horizontal axis is the time and the vertical axis is the logarithm of the temperature.}
\label{fig5}
\end{figure}

\begin{figure}[htp]

  \centering

\begin{tabular}{cc} 
    \includegraphics[width=120mm, height=60mm, keepaspectratio]{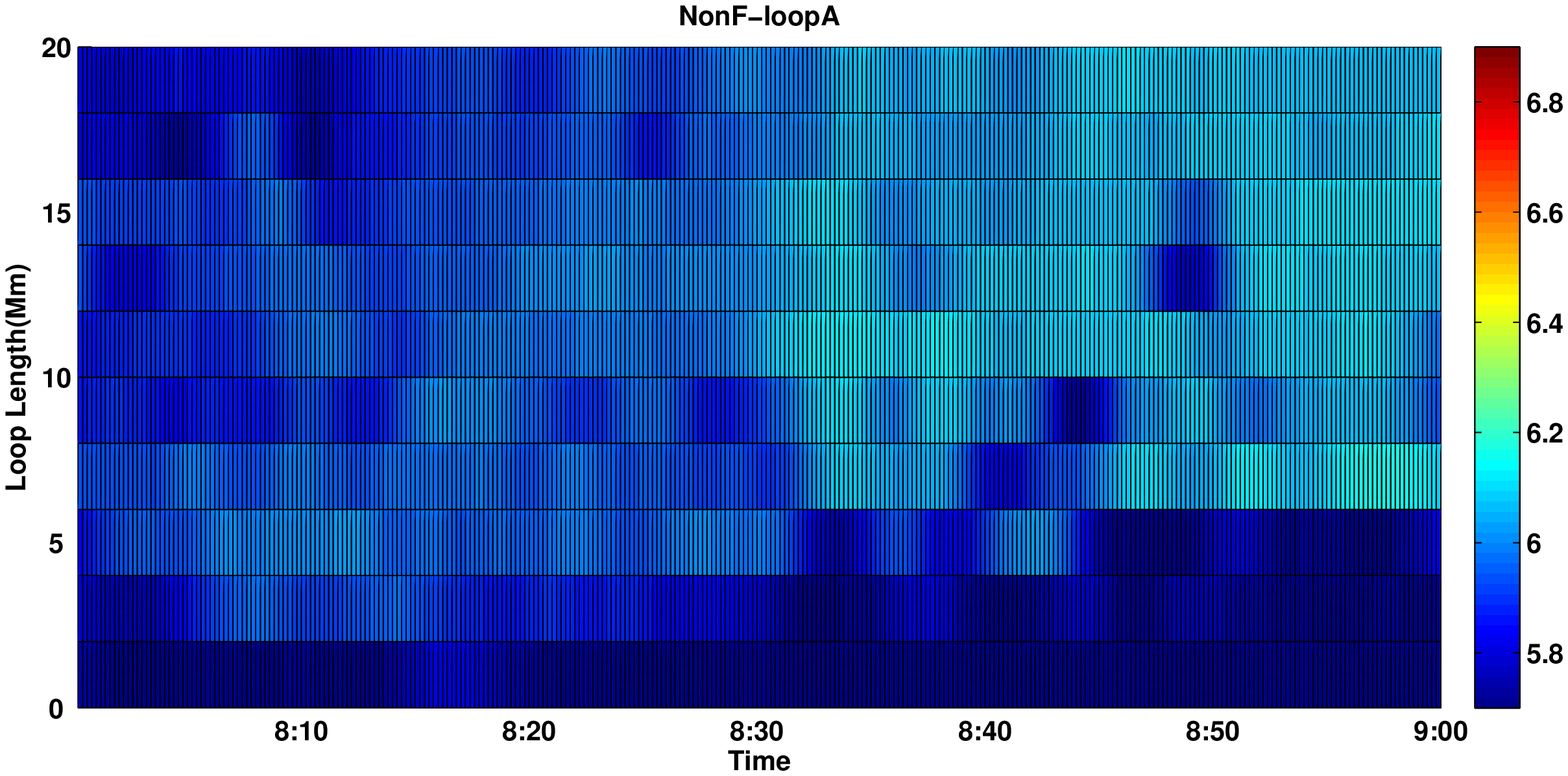}\\
    \includegraphics[width=120mm, height=60mm, keepaspectratio]{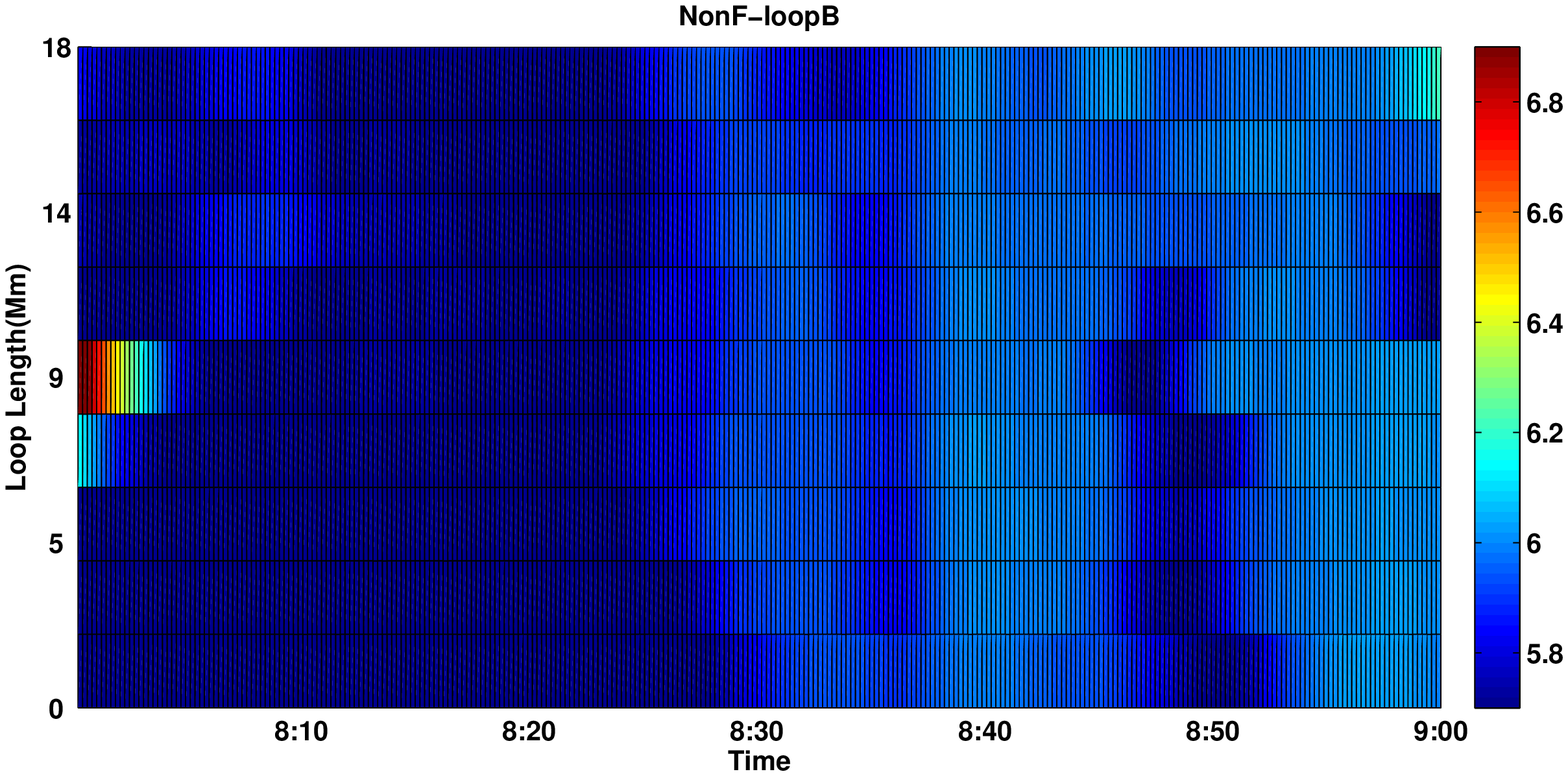}\\
    \includegraphics[width=120mm, height=60mm, keepaspectratio]{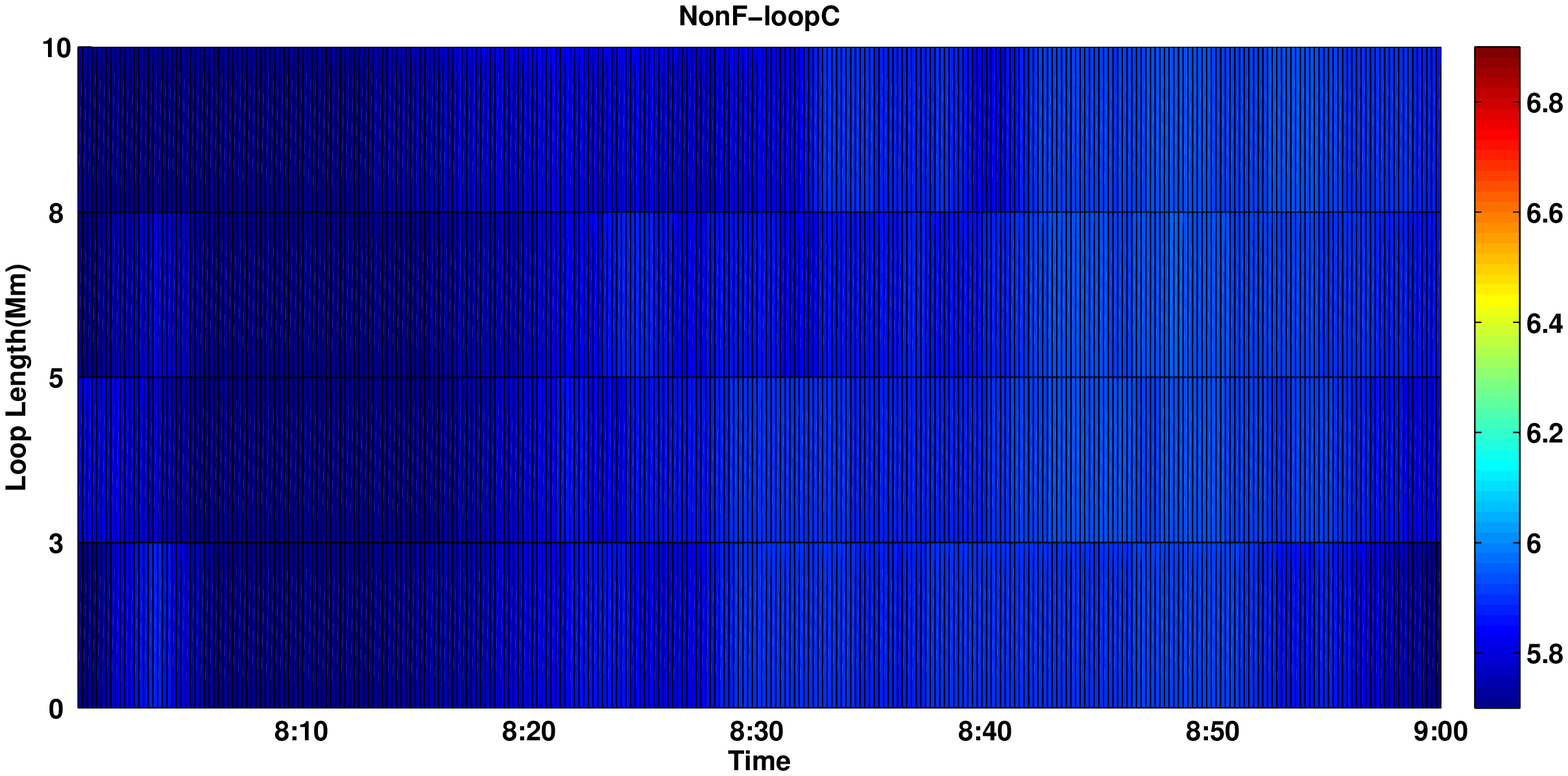}\\
  \end{tabular}
\caption{from top to down: Temperature map of the non-flaring loops A, B and C as a time-series. The vertical axis is the distance along the loop in Mm, and the horizontal axis is the time. The color-bar in the left shows the colors considered for the temperature range.}
\label{fig6}
\end{figure}

% Figure 
%
\begin{figure}
 \centering
 \includegraphics[width=12cm]{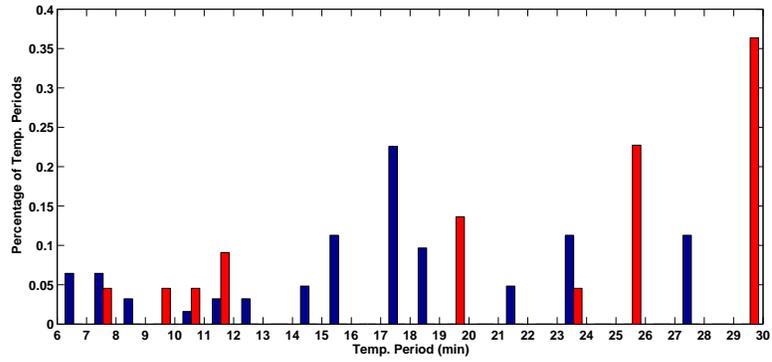}
\caption{Hisogram of the temperature periods percentages for the loops' strips of the flaring (blue bars) and non-flaring (red bars) ARs. The horizontal axis shows the temperature periods in minute.}
\label{fig7}
\end{figure}

\begin{figure}
\centering
\includegraphics[width=12cm]{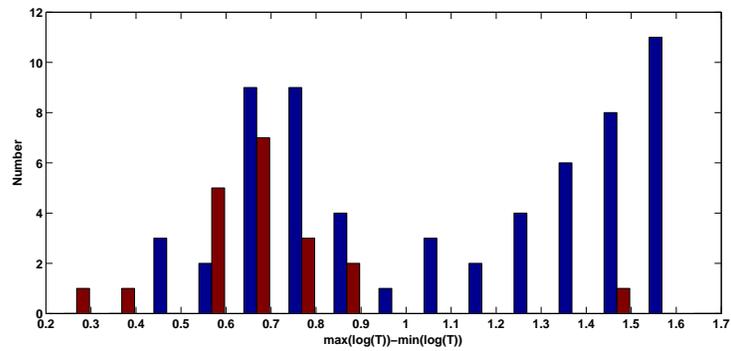}
\caption{Hisogram of the parameter of (max(log(T))-min(log(T))) for each strip of the loops of the flaring (blue bars) and non-flaring (red bars) ARs. 
}
\label{fig8}
\end{figure}

%\begin{figure} 
 %\centering
 %\includegraphics[width=12cm]{table1.eps}
%\caption{Table 1: The properties observed for the loop segments of the flaring AR.}
%\label{table1}
%\end{figure}
%\begin{figure} 
 %\centering
%\includegraphics[width=6cm]{table2.eps}
%\caption{Table 2: The properties observed for the loop segments of the non flaring AR.}
%\label{table2}
%\end{figure}
%
%               one-column-spanning table
%________________________________________ Table 2: Use_of_the routines
\begin{table}[]
\caption[]{The properties observed for the loop segments of the flaring AR.
\label{table1}}
\begin{tabular}{llllll}
\multicolumn{1}{c}{\begin{tabular}[c]{@{}c@{}}FLoopA\\ (Strip Number)\end{tabular}} & \multicolumn{1}{c}{\begin{tabular}[c]{@{}c@{}}The highest \\ Temp.'s period   \\ observed\end{tabular}} & \multicolumn{1}{c}{\begin{tabular}[c]{@{}c@{}}Max(log(T))-\\ Min(log(T))\end{tabular}} & \multicolumn{1}{c}{\begin{tabular}[c]{@{}c@{}}FLoopB2\\ (Strip Number)\end{tabular}} & \multicolumn{1}{c}{\begin{tabular}[c]{@{}c@{}}The highest \\ Temp.'s period   \\ observed\end{tabular}} & \multicolumn{1}{c}{\begin{tabular}[c]{@{}c@{}}Max(log(T))-\\ Min(log(T))\end{tabular}} \\
1                                                                                   & 9.94                                                                                                    & 1.09                                                                                   & 1                                                                                    & 18.07                                                                                                   & 0.68                                                                                   \\
2                                                                                   & 16.57                                                                                                   & 0.79                                                                                   & 2                                                                                    & 24.85                                                                                                   & 0.83                                                                                   \\
3                                                                                   & 8.46                                                                                                    & 0.65                                                                                   & 3                                                                                    & 24.85                                                                                                   & 0.85                                                                                   \\
4                                                                                   & 28.4                                                                                                    & 1.11                                                                                   & 4                                                                                    & 7.36                                                                                                    & 0.84                                                                                   \\
5                                                                                   & 28.4                                                                                                    & 0.75                                                                                   & 5                                                                                    & 8.64                                                                                                    & 0.85                                                                                   \\
6                                                                                   & 24.85                                                                                                   & 0.76                                                                                   & 6                                                                                    & 8.28                                                                                                    & 0.93                                                                                   \\
7                                                                                   & 22.09                                                                                                   & 0.58                                                                                   & 7                                                                                    & 18.07                                                                                                   & 0.84                                                                                   \\
8                                                                                   & 18.07                                                                                                   & 1.55                                                                                   & 8                                                                                    & 28.4                                                                                                    & 0.73                                                                                   \\
9                                                                                   & 18.07                                                                                                   & 1.6                                                                                    & \multicolumn{1}{c}{FLoopC1}                                                          & \multicolumn{1}{c}{-}                                                                                   & \multicolumn{1}{c}{-}                                                                  \\
10                                                                                  & 12.42                                                                                                   & 1.57                                                                                   & 1                                                                                    & 28.4                                                                                                    & 1.46                                                                                   \\
11                                                                                  & 12.42                                                                                                   & 1.42                                                                                   & 2                                                                                    & 22.09                                                                                                   & 1.34                                                                                   \\
12                                                                                  & 24.85                                                                                                   & 1.56                                                                                   & 3                                                                                    & 16.57                                                                                                   & 1.36                                                                                   \\
13                                                                                  & 19.88                                                                                                   & 1.6                                                                                    & 4                                                                                    & 28.04                                                                                                   & 1.49                                                                                   \\
14                                                                                  & 19.88                                                                                                   & 1.24                                                                                   & 5                                                                                    & 24.85                                                                                                   & 1.6                                                                                    \\
15                                                                                  & 18.07                                                                                                   & 1.58                                                                                   & 6                                                                                    & 24.85                                                                                                   & 1.42                                                                                   \\
16                                                                                  & 19.88                                                                                                   & 1.45                                                                                   & 7                                                                                    & 15.29                                                                                                   & 1.6                                                                                    \\
17                                                                                  & 16.57                                                                                                   & 0.7                                                                                    & 8                                                                                    & 13.25                                                                                                   & 1.56                                                                                   \\
18                                                                                  & 7.36                                                                                                    & 1.6                                                                                    & 9                                                                                    & 13.25                                                                                                   & 1.6                                                                                    \\
19                                                                                  & 8.64                                                                                                    & 0.95                                                                                   & 10                                                                                   & 16.57                                                                                                   & 1.6                                                                                    \\
20                                                                                  & 16.57                                                                                                   & 1.54                                                                                   & 11                                                                                   & 16.57                                                                                                   & 1.6                                                                                    \\
21                                                                                  & 7.36                                                                                                    & 1.18                                                                                   & 12                                                                                   & 9.46                                                                                                    & 1.13                                                                                   \\
22                                                                                  & 7.36                                                                                                    & 1.51                                                                                   & \multicolumn{1}{c}{FLoopC1}                                                          & \multicolumn{1}{c}{-}                                                                                   & \multicolumn{1}{c}{-}                                                                  \\
23                                                                                  & 18.07                                                                                                   & 1.58                                                                                   & 1                                                                                    & 18.07                                                                                                   & 0.88                                                                                   \\
24                                                                                  & 22.09                                                                                                   & 1.33                                                                                   & 2                                                                                    & 28.4                                                                                                    & 0.8                                                                                    \\
25                                                                                  & 24.85                                                                                                   & 0.72                                                                                   & 3                                                                                    & 15.29                                                                                                   & 0.87                                                                                   \\
\multicolumn{1}{c}{FLoopB1}                                                         & \multicolumn{1}{c}{-}                                                                                   & \multicolumn{1}{c}{-}                                                                  & 4                                                                                    & 16.57                                                                                                   & 0.93                                                                                   \\
1                                                                                   & 18.07                                                                                                   & 1.43                                                                                   & 5                                                                                    & 18.07                                                                                                   & 1.22                                                                                   \\
2                                                                                   & 15.29                                                                                                   & 0.76                                                                                   & 6                                                                                    & 28.4                                                                                                    & 0.58                                                                                   \\
3                                                                                   & 18.07                                                                                                   & 0.76                                                                                   &                                                                                      &                                                                                                         &                                                                                        \\
4                                                                                   & 18.07                                                                                                   & 0.75                                                                                   &                                                                                      &                                                                                                         &                                                                                        \\
5                                                                                   & 18.07                                                                                                   & 0.59                                                                                   &                                                                                      &                                                                                                         &                                                                                        \\
6                                                                                   & 19.88                                                                                                   & 0.8                                                                                    &                                                                                      &                                                                                                         &                                                                                        \\
7                                                                                   & 19.88                                                                                                   & 0.91                                                                                   &                                                                                      &                                                                                                         &                                                                                        \\
8                                                                                   & 19.88                                                                                                   & 1.36                                                                                   &                                                                                      &                                                                                                         &                                                                                        \\
9                                                                                   & 11.04                                                                                                   & 1.6                                                                                    &                                                                                      &                                                                                                         &                                                                                        \\
10                                                                                  & 18.07                                                                                                   & 1.6                                                                                    &                                                                                      &                                                                                                         &                                                                                        \\
11                                                                                  & 18.07                                                                                                   & 1.6                                                                                    &                                                                                      &                                                                                                         &                                                                                       
\end{tabular}
\end{table}
%%%%%
%%%%%
%%%%
\begin{table}[]
\caption[]{The properties observed for the loop segments of the non flaring AR.\label{table2}}
\begin{tabular}{lll}
\multicolumn{1}{c}{\begin{tabular}[c]{@{}c@{}}Nonf-LoopA\\ (Strip Number)\end{tabular}} & \multicolumn{1}{c}{\begin{tabular}[c]{@{}c@{}}The highest \\ Temp.'s period   \\ observed\end{tabular}} & \multicolumn{1}{c}{\begin{tabular}[c]{@{}c@{}}Max(log(T))-\\ Min(log(T))\end{tabular}} \\
1                                                                                       & 24                                                                                                      & 0.61                                                                                   \\
2                                                                                       & 30                                                                                                      & 0.95                                                                                   \\
3                                                                                       & 30                                                                                                      & 0.81                                                                                   \\
4                                                                                       & 20                                                                                                      & 1.51                                                                                   \\
5                                                                                       & 20                                                                                                      & 0.77                                                                                   \\
6                                                                                       & 20                                                                                                      & 0.81                                                                                   \\
7                                                                                       & 11.42                                                                                                   & 0.71                                                                                   \\
8                                                                                       & 12                                                                                                      & 0.73                                                                                   \\
9                                                                                       & 30                                                                                                      & 0.72                                                                                   \\
10                                                                                      & 30                                                                                                      & 0.77                                                                                   \\
11                                                                                      & 30                                                                                                      & 0.61                                                                                   \\
\multicolumn{1}{c}{\begin{tabular}[c]{@{}c@{}}Nonf-LoopB\\ (Strip Number)\end{tabular}} & \multicolumn{1}{c}{\begin{tabular}[c]{@{}c@{}}The highest \\ Temp.'s period \\ observed\end{tabular}}   & \multicolumn{1}{c}{\begin{tabular}[c]{@{}c@{}}Max(log(T))-\\ Min(log(T))\end{tabular}} \\
1                                                                                       & 26.66                                                                                                   & 0.36                                                                                   \\
2                                                                                       & 26.66                                                                                                   & 0.64                                                                                   \\
3                                                                                       & 10.43                                                                                                   & 0.45                                                                                   \\
4                                                                                       & 12                                                                                                      & 0.62                                                                                   \\
5                                                                                       & 30                                                                                                      & 0.98                                                                                   \\
6                                                                                       & 8.57                                                                                                    & 0.67                                                                                   \\
\multicolumn{1}{c}{\begin{tabular}[c]{@{}c@{}}Nonf-LoopC\\ (Strip Number)\end{tabular}} & \multicolumn{1}{c}{\begin{tabular}[c]{@{}c@{}}The highest \\ Temp.'s period \\ observed\end{tabular}}   & \multicolumn{1}{c}{\begin{tabular}[c]{@{}c@{}}Max(log(T))-\\ Min(log(T))\end{tabular}} \\
1                                                                                       & 26.66                                                                                                   & 0.76                                                                                   \\
2                                                                                       & 26.66                                                                                                   & 0.75                                                                                   \\
3                                                                                       & 26.66                                                                                                   & 0.26                                                                                   \\
4                                                                                       & 30                                                                                                      & 0.27                                                                                   \\
5                                                                                       & 30                                                                                                      & 0.8                                                                                   
\end{tabular}
\end{table}

\section{acknowledgements}
The author Narges Fathalian wishes to also express her thanks for the technical support and comments which has received from Dr.Farhad Daii and Dr.Mohsen Javaherian regarding to this work. 
%\end{acknowledgements}
%
\newpage
\bibliographystyle{raa}
\bibliography{refs}
%\begin{thebibliography}{}
%
%\end{thebibliography}
\end{document}